\documentclass[pdflatex,sn-apa,longnamesfirst,sort,numbers,sort&compress]{sn-jnl}% APA Reference Style 

\usepackage{graphicx}
\usepackage{caption}
\usepackage{subcaption}
\usepackage{tikz}
\usetikzlibrary{positioning, calc}
\usepackage{float}
\usepackage{tabularx}
\usepackage{longtable}
\usepackage{multirow}
\usepackage{amsmath,amssymb,amsfonts}
\usepackage{mathrsfs}
\usepackage[title]{appendix}
\usepackage{xcolor}
\usepackage{textcomp}
\usepackage{manyfoot}
\usepackage{booktabs}
\usepackage{algorithm}
\usepackage{algorithmicx}
\usepackage{algpseudocode}
\usepackage{listings}
\usepackage{lmodern}
\usepackage{lineno}
\usepackage{comment}
\usepackage{setspace}
\usepackage{ragged2e}
\usepackage{placeins} % Para \FloatBarrier

\newcommand{\indep}{\perp \!\!\! \perp}
\newcommand{\noindep}{\not\!\perp\!\!\!\perp}
\usepackage{hyperref}
\hypersetup{
  colorlinks=true,
  linkcolor=blue,
  filecolor=magenta,      
  urlcolor=cyan,
}

\begin{document}
\sloppy

\title[SEMRBS]{Integration of Structural Equation Modeling and Bayesian Networks in the Context of Causal Inference: A Case Study on Personal Positive Youth Development}

\author*[1,2]{\fnm{Edgar} \sur{Benitez}}\email{ebenitezs@unav.es}
\author[3]{\fnm{Álvaro} \sur{Balaguer}}\email{abalaguer@unav.es}

\affil*[1]{\orgdiv{Institute for Data Science and Artificial Intelligence}, \orgname{University of Navarra}, \orgaddress{\street{Campus Universitario}, \city{Pamplona}, \postcode{31008}, \state{Navarra}, \country{Spain}}}
\affil[2]{\orgdiv{Tecnun Escuela de Ingeniería}, \orgname{University of Navarra}, \orgaddress{\street{P° de Manuel Lardizabal, 13}, \city{San Sebastian}, \postcode{20018}, \state{Gipuzkoa}, \country{Spain}}}
\affil[3]{\orgdiv{School of Education and Psychology}, \orgname{University of Navarra}, \orgaddress{\city{Pamplona}, \postcode{31008}, \state{Navarra}, \country{Spain}}}
\abstract{In this study, the combined use of structural equation modeling (SEM) and Bayesian network modeling (BNM) in causal inference analysis is revisited. The perspective highlights the debate between proponents of using BNM as either an exploratory phase or even as the sole phase in the definition of structural models, and those advocating for SEM as the superior alternative for exploratory analysis. The individual strengths and limitations of SEM and BNM are recognized, but this exploration evaluates the contention between utilizing SEM's robust structural inference capabilities and the dynamic probabilistic modeling offered by BNM. A case study of the work of, \citet{balaguer_2022} in a structural model for personal positive youth development (\textit{PYD}) as a function of positive parenting (\textit{PP}) and perception of the climate and functioning of the school (\textit{CFS}) is presented. The paper at last presents a clear stance on the analytical primacy of SEM in exploratory causal analysis, while acknowledging the potential of BNM in subsequent phases.}

\keywords{Causal Inference, Structural Equation Modeling, Bayesian Networks, Youth Development, Positive Psychology, Exploratory Analysis, Educational Outcomes, Predictive Modeling}

%%\pacs[JEL Classification]{D8, H51}

%%\pacs[MSC Classification]{35A01, 65L10, 65L12, 65L20, 65L70}

\maketitle
\section{Introduction}
The pursuit of the "best" algorithm, primarily focused on achieving a better fit, irrespective of its logical coherence, \citet{pearl_2019, darwiche_2017}, has been the cornerstone of recent developments in what can broadly be termed as machine learning (ML) technology. This approach parallels inductive models whose primary concern lies in being verified, yet unfortunately not falsified, \citet{reizinger_2023}. By adhering to Popper’s argumentation, one can argue that if the objective is to verify the hypothesis under examination rather than to falsify it, this approach veers more towards dogmatism than science \cite[p. 51]{popper_1993}. In this context, in recent years, serious ethical challenges have emerged in the use of ML, as pointed out by Zhang, \citet{zhang_2023}, which underscores how the unidimensional focus on algorithmic efficacy can overlook critical considerations regarding its social and moral impact. This emphasis on predictive optimization over transparency and accountability amplifies the risk of adopting a dogmatic approach, limiting the field's capacity for self-correction and ethical evolution.

One cannot ignore that the results generated by ML are evident and of great utility, but they can be shown to be merely operationalizations of already achieved knowledge, with Pearl seeing in them ``purely statistical relationships defined by the naked data'', \citet{pearl_2019}. This approach to addressing the problem of knowledge acquisition falls within what is known as an inductive process, for which, although the evidence may statistically increase the probability of a hypothesis, this increase does not constitute genuine support. The evidence can counteract parts of the hypothesis that are not directly implicated by it, \citet{popper_1983}., \citet{popper_1983} propose different theorems within the epistemological context that emphasize:

\textbf{Theorem 1:} If the probability of the hypothesis given the evidence, \( p(h, e) \), is not equal to the probability of the evidence, \( p(e) \), then the probability of the hypothesis given the evidence is less than the probability of the hypothesis implied by the evidence:
\[ p(h \leftarrow e, e) < p(h \leftarrow e) \]
This shows that the evidence \( e \) counteracts the part of the hypothesis \( h \) that is not directly deduced from \( e \).

\textbf{Theorem 2:} Under the same assumptions, the difference in the probability of the hypothesis given the evidence and the probability of the hypothesis implied by the evidence is equal to the excess of the probability of their conditional over the conditional probability:
\[ p(h \leftarrow e) - p(h \leftarrow e, e) = p(\neg h, e) p(\neg e) \]
This excess \( p(\neg h, e) p(\neg e) \) is positive, demonstrating that \( e \) strongly countersupports \( h \) more than what is typically accounted for by \( p(h, e) \).

Thus, while statistical methods can provide operational utility, they often fail to offer the kind of inductive support required for genuine scientific advancement, as will be further demonstrated later in this text by, \citet{pearl_2009}.

In this regard, if there is an aspiration to adopt a \textit{demarcated}\footnote{Demarcation: A criterion upon which the distinction between scientific and non-scientific statements is based, emphasizing the susceptibility of scientific theories to be falsified, \citet{popper_1993}.} approach from the scientific perspective, it is necessary to work on hypotheses that seek the falsification of hypotheses. In this context, the theory of causal inference, along with its derivatives: association, intervention, and counterfactual strategies, based on \textit{structural}\footnote{Structural: The researcher incorporates causal assumptions as part of the model. Each equation is a representation of causal relationships between a set of variables, and the form of each equation conveys the assumptions that the analyst has asserted. \cite[p.~304]{bollen_2013}.} models, becomes indispensable, \citet{pearl_2019, kasirzadeh_2021, carey_2022}.

However, any attempt to propose structural-inferential models must start from information captured by some type of experiment. For this purpose, the literature recognizes two types, \textit{manipulative experiments} and \textit{mensurative experiments}, \citet{hurlbert_1984}. \textit{Manipulative experiments}, those in which the researcher actively manipulates one or more independent variables, to date this is the only methodology that strictly satisfies the identification of causal dependence, provided that they are based on the assumption that controls are used, replications are present, the assignment of the independent variable is random, and the estimation of experimental error is unbiased, \citet{milliken_1984, hurlbert_1984}. For all other cases, i.e., \textit{mensurative experiments}---those that measure variables without manipulation, \citet{hurlbert_1984}---alternatives for validating causality are based on theoretical \textit{structural} assumptions, \citet{pearl_2009, rubin_1974} or, in general, on hypothetico-deductive approaches, \citet{popper_1993}.

For many years, one of the methodologies for evaluating \textit{structural} models was SEM. However, over time, many authors began to issue judgments arguing that this methodology lacked the capacity to assess inference models, \citet{guttman_1977, freedman_1981, baumrind_1983, cliff_1983, freedman_1987, goldthorpe_2001, freedman_2004}, initiating, in Pearl's words, one of the most ``\textit{bizarre}'' stages in the history of science:

\begin{quote}
\textit{"We are witnessing one of the most bizarre circles in the history of science: causality in search of a language and, simultaneously, speakers of that language in search of its meaning."}, \citet{pearl_2009}
\end{quote}

\begin{figure}[!ht]
\centering
\begin{tikzpicture}
\node (x) at (0,0) {Cause (X)};
\node (y) at (6,0) {Consequence (Y)};
\node (U_X) at (0,-2) {Factors affecting X \((U_X)\)};
\node (U_Y) at (6,-2) {Factors affecting Y \((U_Y)\)};

\draw[->] (x) -- (y) node[midway,above] {\(\beta\) (Path coefficient)};
\draw[->,dashed] (U_X) -- (x);
\draw[->,dashed] (U_Y) -- (y);
\end{tikzpicture}
\caption{The path diagram showing the absence of a causal path from \( Y \) to \( X \)}
\label{fig:path-diagram}
\end{figure}
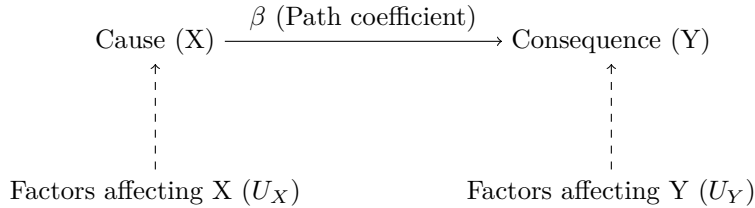

Pearl himself demonstrates that this confusion is related to conflating causality with algebraic equations in SEM, \citet{pearl_2009}, which had already been warned by, \citet{popper_1983} at the time. This is explained by the symmetry of algebraic expressions, summarized in that equations themselves do not imply causality; for instance, the equation \(y = \beta x + u_Y\) can be algebraically manipulated to \(x = (y - u_Y) / \beta\), which might wrongly suggest that \(y\) causes \(x\). In his time, Sewall Wright, \citet{wright_1921} observed this problem and introduced the concept of path diagrams that represent the direction of causality and differentiate it from mere algebraic association. The path diagram, Figure \ref{fig:path-diagram}, explicitly shows the absence of a causal path from \(Y\) to \(X\).

In this formulation, \(\beta\) is the path coefficient quantifying the direct causal effect of the cause on the consequence, and the dashed arrows represent the contribution of unobserved exogenous variables. The variables \(U_X\) and \(U_Y\) are considered \textit{exogenous}; they represent observed or unobserved background factors that the modeler decides not to explain, meaning they are factors that influence but are not influenced by the other variables (called \textit{endogenous}) in the model. Unobserved exogenous variables represent factors omitted from the model but considered relevant for explaining the behavior of the variables within it. Thus, background factors in structural equations differ fundamentally from residual terms in regression equations. The latter are artifacts of analysis which, by definition, are uncorrelated with the regressors. The former are actual and responsible for the observed variations in the data. This difficulty in distinguishing the \textit{structural} model from the regression model may explain why the aforementioned authors have declared that SEM analysis lacks the capacity to evaluate causality models.

The consequences are still observed in work approaches, as many researchers relegate the ability of SEM to evaluate causal models, falling into even more \textit{bizarre} situations where other methodologies that are openly not causal but associative like BNM are given roles of causal inference, \citet{zheng_2009}.

This paper seeks to demonstrate an approach adjusted to the causal inference functionalities that can be attributed to SEM and how, through its capacity to establish causal relationships in data, it can serve as a basis for using BNM. This can overcome some of the difficulties inherent in SEM, such as the non-linearity of the relationships between variables, and also expand its inferential capacity to what Pearl has proposed under the approaches of second and third-level causality: \textit{intervention} and \textit{counterfactuals}, \citet{pearl_2009, pearl_2019}.

\section{Structural Equation Modeling (SEM)}
The use of randomized experiments, mentioned in previous paragraphs, only became a widespread practice from the second half of the 19th century, \citet{rubin_1974}. Therefore, if it were considered that the only way to evaluate causality relationships was through this type of experiment, progress in many areas of science would be impossible. This is because the inability to control the assignment of factors ranges from ethical motivations to simple logistical problems (e.g., randomizing a person's gender or socioeconomic level), \citet{rubin_1974}. It is in this context that techniques such as SEM or non-randomized experiments become indispensable for their capacity to evaluate and estimate causal effects.

Specifically, SEM was developed to enable observational studies to make advancements in solving causal hypotheses, \citet{wright_1921, pearl_2009}. It offers increased flexibility in model specifications, including simultaneous and nonlinear equations (there are specific cases of applications to non-linear models in variables but linear in parameters, \citet{bollen_2013}), within a structural framework called \textit{path analysis}. It allows for comprehensive modeling of complex and multiple causal relationships among variables, including mediating and moderating effects, which operationalizes the decomposition of direct and indirect effects in causal relationships, and facilitates the incorporation of latent variables. SEM provides metrics for assessing model fit and includes advanced techniques for handling missing data without the need for imputation. Furthermore, this framework enables the integration of existing theories for the direct evaluation of complex theories and allows for comparisons between different theoretical models to determine the best fit to the observed data. For a theory of causality, it employs a robust mathematical language to represent causal questions.

The only major restriction on the use of SEM is understood not from an instrumental viewpoint but from the relevance of its \textit{structural} formulation, implying that if one intends to use SEM, the researcher must first have knowledge of the causality of the system. With this, the discussion is initiated that a model which does not preliminarily take into account the researcher's knowledge cannot be considered causal. In his early works, J{\"o}reskog made this clear:

\begin{quote}
``A typical application of the procedure [SEM] is in confirmatory factor studies, where the experimenter has already obtained a certain amount of knowledge about the variables measured and is therefore in a position to formulate a hypothesis that specifies some of the factors involved." [Note: SEM refers to Structural Equation Modeling].\cite[p.~183]{joreskog_1969}
\end{quote}

The model-building process of a SEM involves analyzing two distinct types of models: a confirmatory measurement (factor analysis) model that outlines the relationship between observed measures and their underlying constructs, allowing these constructs to intercorrelate freely; and a confirmatory structural (SEM) model that defines the causal relationships between the constructs as proposed by a specific theory, \citet{anderson_1982,joreskog_1984}.

Utilizing the LISREL program notation, the confirmatory factor analysis model, also known as the confirmatory measurement model as proposed by J{\"o}reskog and Sörbom, \citet{joreskog_1984}, is articulated as:

\begin{equation}
    x = \Lambda\xi + \delta,
\end{equation}
where \( x \) represents a vector of \( q \) observed measures, \( \xi \) is a vector of \( n \) underlying factors with \( n < q \), \( \Lambda \) denotes a \( q \times n \) matrix of pattern coefficients or factor loadings that link the observed measures to the underlying construct factors, and \( \delta \) is a vector of \( q \) variables symbolizing random measurement error and measure specificity. It is postulated within this model that \( E(\xi\delta') = 0 \). The variance-covariance matrix for \( x \), designated as \( \Sigma \), is defined by the equation:
\begin{equation}
    \Sigma = \Lambda\Phi\Lambda' + \Theta_\delta,
\end{equation}
with \( \Phi \) being the \( n \times n \) covariance matrix of \( \xi \) and \( \Theta_\delta \) representing the diagonal \( q \times q \) covariance matrix of \( \delta \).

Similarly, the confirmatory structural model that embodies the relationships among endogenous and exogenous constructs is given by:
\begin{equation}
    \eta = B\eta + \Gamma\xi + \zeta,
\end{equation}
where \( \eta \) is a vector of \( m \) endogenous constructs, \( \xi \) encapsulates a vector of \( n \) exogenous constructs, \( B \) is an \( m \times m \) matrix detailing the effects of the endogenous constructs on one another, \( \Gamma \) is an \( m \times n \) matrix describing the influence of the exogenous constructs on the endogenous constructs, and \( \zeta \) is a vector of \( m \) residuals indicative of errors in equations and random disturbance terms.

It is generally accepted that these analyses---measurement and structural---represent extremes of a continuum that, for operationalization reasons, are discretized, \citet{horn_1965,lloret_2014}. That is, the structure and, therefore, causality are defined \textit{a priori} by the researcher. This approach clearly differentiates the descriptive stage (conceptual structure) from the inferential stage (structural equations). Nonetheless, both new and old proposals have been presented that seek to integrate these two stages into either a single-step or a two-step process., \citet{anderson_1988} show that the one-step modeling approach presents several challenges when compared to a two-step approach. First, \textit{interpretational confounding} can occur, where the empirical meaning assigned to unobserved variables deviates from their theoretical meaning due to model parameter specification. This leads to significant changes in the interpretation of theoretical constructs. Second, coefficient estimates can change markedly when alternative structural models are estimated, suggesting that the meaning of estimated constructs may vary with different model specifications, complicating the accurate interpretation of variable relationships. Third, there's a tendency to maximize model fit at the expense of meaningful interpretability of constructs, potentially leading to misleading or incorrect understandings of the relationships between latent and observed variables. Finally, estimating measurement and structural models simultaneously can obscure the source of specification problems, making it difficult to determine whether issues arise from the measurement model, the structural model, or their interaction, potentially resulting in the acceptance of poorly specified models.

Later it will be seen how the argumentation by, \citet{anderson_1988} has gone unnoticed by many practitioners of the BNM approach, seemingly unaware of the challenges: \textit{interpretational confounding}, coefficient estimates, and maximizing model fit. These practitioners often relegate SEM to marginal positions or even to complete disregard.

\section{Bayesian Network Modeling (BNM)}
BNM, also known as Belief Networks or Bayes Nets, are graphical models that use graph theory and conditional probability to represent probabilistic relationships among a set of variables, \citet{pearl_1988, pearl_2009}. Formally, it is defined as a directed acyclic graph (DAG) \(G = (V, E)\), where \(V\) is a set of nodes corresponding to the random variables \(X_1, X_2, \ldots, X_n\) and \(E\) is a set of directed edges (arrows with direction) between these nodes (parents to children). The parents of a variable represent its direct causes, while its ancestors encompass both its direct and indirect causes. Furthermore, a variable's children are those variables it directly causes, and its descendants are all variables that are either directly or indirectly caused by it, \citet{kline_1999}. This model correspond to a Markovian model, where the state of the children depends only on the present state and not on the sequence of their ancestors. A key theorem in this context is the \textit{Causal Markov Condition}, which states that any distribution generated by a Markovian model can be factorized based on the structure of the causal diagram, that is, the joint probability distribution over the variables can be decomposed into a product of conditional probabilities, given by:

\begin{equation}
P(X_1, X_2, \ldots, X_n) = \prod_{i=1}^{n} P(X_i | Pa(X_i))
\end{equation}

where \(Pa(X_i)\) denotes the set of parent nodes of \(X_i\) in the graph \(G\), representing the variables on which \(X_i\) is conditionally dependent (\(\noindep\)). The conditional independence (\(\indep\)) properties of the BNM are determined by its graphical structure. If a variable \(X\) is conditionally independent of its non-descendants given its parents in the network, then the network structure can be used to simplify the computation of the joint probability distribution.

Exact inference in BNM is known to be an NP-hard problem, which means that solving problems of exact inference in BNM may require exponential time relative to the size of the network. To solve this problem, Judea Pearl, \citet{pearl_1988} introduced efficient approaches for approximate inference, such as belief propagation.

\begin{figure}[ht]
\centering
\begin{tikzpicture}[node distance=1cm and 0.5cm]

% Chain structure
\node (X) {X};
\node[right=of X] (W) {W};
\node[right=of W] (Y) {Y};

\draw[->] (X) -- (W);
\draw[->] (W) -- (Y);

\node[align=center, below] at ($(W) - (0,1)$) {Simple\\ \( X \noindep Y \) \\ \( X \indep Y \,|\, W \)};

% Fork structure
\node[right=2cm of Y] (X2) {X};
\node[right=of X2] (W2) {W};
\node[right=of W2] (Y2) {Y};

\draw[<-] (X2) -- (W2);
\draw[->] (W2) -- (Y2);

\node[align=center, below] at ($(W2) - (0,1)$) {Confounder\\ \( X \noindep Y \) \\ \( X \indep Y \,|\, W \)};

% Collider structure
\node[right=2cm of Y2] (X3) {X};
\node[right=of X3] (W3) {W};
\node[right=of W3] (Y3) {Y};

\draw[->] (X3) -- (W3);
\draw[->] (Y3) -- (W3);

\node[align=center, below] at ($(W3) - (0,1)$) {Collider\\ \( X \indep Y \) \\ \( X \noindep Y \,|\, W \)};

\end{tikzpicture}
\caption{Graphical structures in Bayesian networks and their conditional dependencies and independencies.}
\label{fig:BN-structures}
\end{figure}
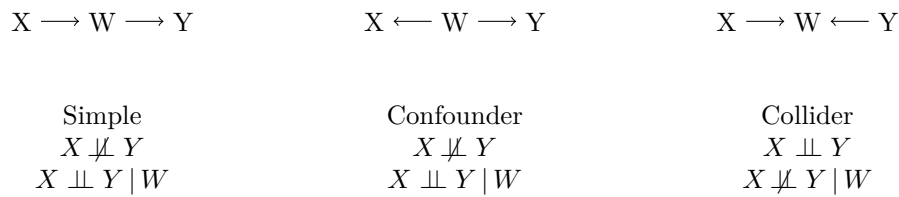

BNM have three basic structures (Markovian chains) which define the relationships between variables, as shown in Figure \ref{fig:BN-structures}. In the simple chain, if W is not considered, X and Y are dependent, but if conditioned on W, X and Y are independent. For confounder, something similar occurs, and for collider, if W is not considered, X and Y are conditionally independent, but if W is considered, X and Y will be conditionally dependent.

\FloatBarrier

The origins of BNM can be traced back to the fields of artificial intelligence and statistics in the latter half of the 20th century. Initially, this methodology was developed to address ML issues, where it has been employed to model the joint probability of a set of variables. This provides a framework for representing uncertain knowledge and dependencies among variables. Thus, BNM serve as versatile tools for tackling problems that involve complex variable dependencies, uncertainty, and the need for probabilistic inference in ML applications. Being a Markovian model characterized by its acyclic structure and the joint independence of error terms, BNM plays a crucial role in evaluate causal inference, as they provide a framework for representing complex causal relationships in a structured and interpretable manner. This integration allows for the dynamic updating of beliefs regarding the causal structure upon the observation of new data, positioning BNM as a powerful tool for validating pre-existing causal theories, albeit often utilized in exploratory analyses.

\section{BNM Structure Learning}
Upon reaching this section, it is necessary to demonstrate the semantic inconsistencies of what a BNM represents:
\begin{quote}
"A Bayesian network is a representational device that is meant to organize one’s knowledge about a particular situation into a coherent whole.", \citet{darwiche_2009}
\end{quote}

This brings us back to the epistemological discussion of what is understood by the word 'knowledge', not just the word but what it represents and therefore the way it is obtained.
Nevertheless, the search for learning structures from 'naked-data' inductive processes has prospered with the development of different algorithms that base their success on fitting data to probabilistically constructed models. Essentially, the process of learning the structure of the BNM involves two methodologies: constraint-based and score-based approaches, \citet{glymour_2019, huang_2018}.
Constraint-based search algorithms aim to efficiently search for a Markov Equivalence Class (MEC)\footnote{Markov Equivalence Class (MEC): a collection of all directed acyclic graphs with the same d-separation, which implies that two nodes are separated by a set of intermediate nodes, considering that all paths between the nodes are blocked by that set.} of graphs that most faithfully represent the observed conditional independence relationships. The goal is to find the causal structure that best fits these conditional independencies without contradicting them. Score-based search algorithms estimate the conditional dependencies or independencies of each variable with respect to independent noise. It uses this information to construct a directed graphical model. Instead of focusing directly on conditional independencies, these algorithms seek to maximize a scoring function that evaluates how well the proposed model fits the observed data.

All in all, these methods are based on the statistical properties of the data, referred to as 'data-driven' structure learning. This approach leaves little room for theorists in a field to incorporate their knowledge into the model's priors.

Faithful to the impossibility of inductive probability, \citet{popper_1983, popper_1993}, this article presents a third category, a `theory-driven' structure learning. This method uniquely leverages the causal inference framework of SEM to inform the BNM's structure, integrating empirical data with established theoretical insights. Unlike the data-driven focus of constraint and score-based methods, the theory-driven approach endows the BNM with an architecture built on a complete and validated \textit{prior} knowledge of causal relationships derived from SEM analyses. This approach not only enriches BNM with theoretically grounded insights but also ensures that the resulting models are closely aligned with substantive domain theories, offering a comprehensive strategy where theoretical knowledge and empirical data converge to define the network structure.

\section{BNM and SEM}
It is intuitively observed that similar approaches are shared by both SEM and BNM. Instead of being considered to generate conflict regarding the use of one against the other, an increasingly wider use of their simultaneous application is being observed: in Medicine and Public Health, \citet{duarte_2011}, in Economics and Social Sciences, \citet{zheng_2009}, in Engineering and Technology, \citet{gupta_2008}, in Transportation and Logistics, \citet{diezmesa_2018, mandhani_2020}, or in Environmental Studies, \citet{xu_2016}, among many others.

However, as expected, the \textit{``bizarre"} era mentioned by Pearl has left its mark on methodological approaches, which can be divided into two groups: those that rely on SEM for proposing causal relationships in exploratory analysis and those that do not. Thus, it should be proposed that if the focus is on causal inference, using BNM in exploratory phases is conceptually and theoretically contradictory. For Pearl, the difficulties of this approach were clear:

\begin{quote}
``In Bayesian analysis, the sensitivity to 'priors' or initial assumptions generally decreases as the sample size increases. However, this is not true for causal assumptions. The sensitivity to prior causal assumptions, such as 'treatment does not change gender,' remains significant regardless of the sample size", \citet{pearl_2009}.
\end{quote}

Therefore, the assignment of values without any conceptual clarity, solely to allow the BNM algorithm to start its iterations in the hope that they will converge to the true value, carries a high risk of bias. In fact, initial works already report that the use of priors far from the true structure led to very high cross-entropy values, and even more so with increases in the equivalent sample size, \citet{heckerman_1995}.

In the group that favors BNM starting the exploration process, works by , \citet{zheng_2009},, \citet{duarte_2011},, \citet{mandhani_2020}, and, \citet{diezmesa_2018} are included. It is worth mentioning that the problems are not limited to the use of BNM in exploratory analysis but even factor analysis lacks rigor, \citet{lloret_2014}, which has been widely reported in the social sciences where it has been the standard methodology for variable evaluation over the last decades.

Although the goal was not an exhaustive review of the literature on the topic, after inspecting the available literature, it was observed that the first approach is the most utilized. Despite the scarcity of examples supporting the alternative hypothesis, \citet{xu_2016, gupta_2008}, these few studies presented address this issue by explaining why they opt for the approach in which the SEM model should provide the causal structure to the BNM. It is asserted that the ability of BNM to incorporate new information for model updating is a well-recognized advantage. However, inaccuracies in the foundational structure of the model may not be corrected with the introduction of new data; inaccuracies may be perpetuated or even exacerbated. In contrast, SEM offers greater flexibility, allowing structural adjustments based on theoretical justification and probabilistic reasoning. This adaptability is crucial for the model's applicability and relevance in various contexts.

Among the criticisms directed at SEM, one in particular is highlighted for its relevance: the difficulty of these models to evaluate nonlinear relationships, \citet{bollen_2013}. Although BNM also do not directly address nonlinear models, the methodology they use to construct networks of conditional probabilities allows the levels of the factors to be examined individually. This facilitates the identification of complex patterns without being restricted to any specific model.

Another limitation attributed to SEM models is their inability to perform model diagnostics, \citet{anderson_2004}, that is, to assess the impact of individual variables within the overall model context. However, this limitation might be more a reflection of the prevailing research approach than an inherent weakness of the SEM models. Since latent variables imply that observed variables are manifestations of broader constructs, any attempt to modify an individual observed variable might be futile without considering its relationship with the corresponding latent variable. Despite this, it is understandable that many researchers continue to evaluate models in a univariate manner, as this approach has historically been the most frequently used.

Lastly, under the causal inference approach, SEM models lack some properties, cannot evaluate intervention effects, counterfactual scenarios, and probabilistic relationships in a more general and flexible manner. In terms of intervention effects, SEM can be adjusted for this purpose but is not specifically designed for it. It is generally more suited for understanding structural relationships in observational data. SEM does not focus on counterfactual questions. Furthermore, although SEM can incorporate measurement errors and other sources of variability, its main focus is on modeling structural relationships rather than probabilistic ones.

Another difference is the idea that the BNM model can be improved with SEM, rather than thinking that they are complementary models that allow achieving the same goal of providing a better explanation of the underlying causal model, allowing the evaluation of non-linear relationships of the variables involved, or allowing the evaluation of the individual performance of explicit variables on the response variable, without blurring their effect into a globalizing latent variable.

\section{Latent and Explicit Variables}
Although the concept of latent variables is powerful in its ability to estimate unobservable effects and recognize complete theoretical constructs, it contrasts with traditional explicit approaches, such as \textit{path analysis} \cite[p.~108]{hatcher_2013}, where each variable is assessed independently with its own identity, the application of BNM predominantly occurs in contexts where explicit variables are emphasized.

However, integrating hidden or latent nodes within BNM offers a unique advantage. This approach not only compiles results but also enables the explicit evaluation of the impact of each indicator within the complete network, thereby providing a comprehensive analysis that leverages the strengths of both latent and explicit variables. Such a methodology allows for a more nuanced understanding of complex phenomena, bridging the gap between observable indicators and underlying processes that are not directly measurable. 

This distinction is particularly evident in fields like medicine, where advancements have significantly relied on the use of explicit biomarkers, essentially following path models, for instance, in clinical research, work on factorial models, \citet{bullen_2021} could make use of these structural models with latent variables alongside explicit biomarkers in a Bayesian Network Model (BNM) could unveil deeper insights into disease mechanisms, patient outcomes, or treatment efficacy. This dual approach facilitates a more robust analysis, capturing both the measured effects of explicit variables and the inferred influences of latent factors. It enriches the model’s explanatory power without sacrificing the detailed evaluation of explicit indicators.

In the words of other researchers, the causal network can be diagnosed, \citet{anderson_2004}. This statement underscores the potential of BNM with latent nodes to diagnose and interpret causal relationships in a way that traditional models, focusing solely on explicit variables, might not achieve. The inclusion of latent variables enables researchers to model and infer complex interactions and causalities that reflect the reality of the phenomena under study more accurately.

Additionally, the use of latent variables can be applied to different uses:

\begin{enumerate}
    \item Handling of missing data: they can treat incomplete data by inferring the most probable values for the missing variables.
    \item Abstraction: they can capture more abstract underlying structures or generative factors in the data, providing insights into complex phenomena.
    \item Simplification: they reduce the model's complexity by grouping several observed variables under a more general concept, making the model more interpretable.
    \item Incorporation of prior knowledge: previously established theoretical relationships can be incorporated into the model through latent variables, enhancing the model's accuracy and relevance.
\end{enumerate}

In recent years, the focus on latent spaces or dimensions has become an important area of work in ML methodologies, \citet{schoelkopf_2021, ahuja_2023, lotfollahi_2023}. This is because there has come a point where ML algorithms are being asked to explain their results, and one of the approaches being intensely pursued is causal models using latent dimensions or spaces. As Popper already predicted, being inductive approaches, these authors arrive at the same conclusions: the results ``are not directly interpretable,'', \citet{lotfollahi_2023}.
In conclusion, the complementary use of latent and explicit variables in BNM opens new avenues for research and analysis across various fields. By embracing both, researchers can construct more comprehensive models that capture the full spectrum of influences on the phenomena of interest, from the directly observable to the implicitly underlying.

\section{Case Study}
\subsection{Positive Youth Development, Parenting and School Climate Study \texorpdfstring{\citet{balaguer_2022}}{}}
This study tested an empirical model of the relationship between \textit{PYD} and two contextual factors: \textit{PP} and \textit{CFS}. The hypothesis tested was that a perception of positive parenting or positive relationship with parents and a positive perception of the climate and functioning of the school will contribute to the prediction of \textit{PYD}. The sample was composed of 1507 adolescents recruited in seven Spanish schools, who were aged between 12 and 18 years; 52\% were female. The structural model evaluated for \textit{PYD} in function of \textit{PP} and \textit{CFS} was made in this way: constructs associated to \textit{PP}: affect and communication (\textit{AfC}), autonomy granting (\textit{Aut}), humor (\textit{Hum}), and self-disclosure (\textit{Dis}); constructs associated to \textit{CFS}: 1) peer relations (\textit{Pee}); 2) school bonds (\textit{Bon}): belongingness (\textit{Bel}) and support (\textit{Sup}); 3) activities proffered (\textit{Pro}); and 4) clarity of rules and values (\textit{Cla}): rules (\textit{Rul}) and values (\textit{Val}); and constructs associated to \textit{PYD}: optimism (\textit{Opt}), pessimism (\textit{Pes}), general self-efficacy (GSe), agency (\textit{Age}), comprehensibility (\textit{Com}), manageability (\textit{Man}), and meaningfulness (\textit{Mea}), Figure \ref{fig:sem_figure}.
Previous reliability and validity analyses of the constructs were carried out, and correlational analyses and structural predictions were made. The results show that both \textit{PP} and \textit{CFS} were associated with better scores in \textit{PYD}, Table \ref{table:model_fit_metrics}.

\subsection{Methodology}
A key component of the methodology was the application of SEM in conjunction with BNM analysis to dissect the interrelations among psychological and behavioral variables in youth development.

\begin{figure}[ht]
    \centering
    \includegraphics[width=\textwidth]{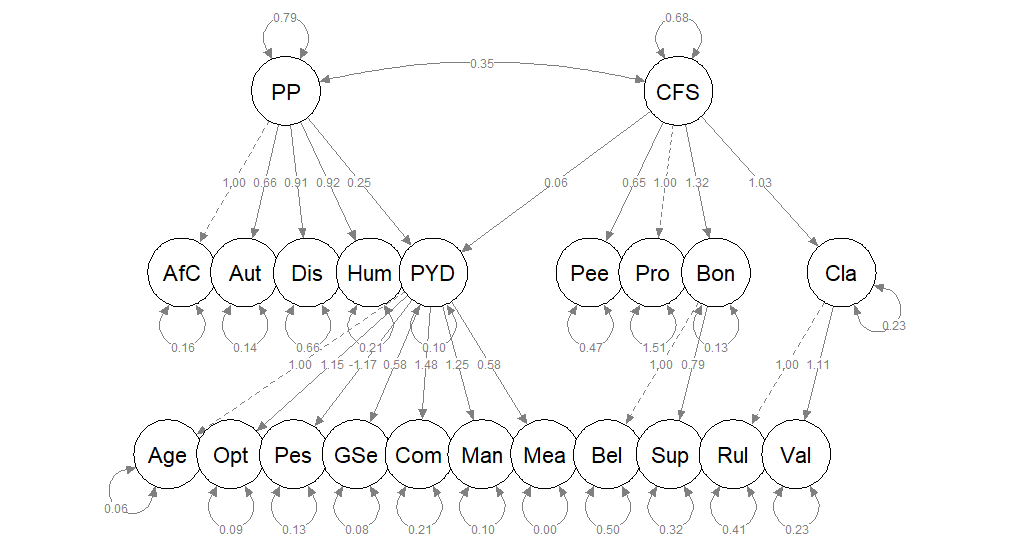}
    \caption{
        Structural model for \textit{PYD} (Personal Positive Youth Development) in function of \textit{PP} (Positive Parenting) and \textit{CFS} (Perception of the Climate and Functioning of the School). 
        \textbf{Note}: Constructs associated to \textit{PP}: Affect and communication (\textit{AfC}), Autonomy (\textit{Aut}), Humor (\textit{Hum}), and Self-disclosure (\textit{Dis}). 
        Constructs associated to \textit{CFS}: 1) Peers relations (\textit{Pee}); 2) Bonds (\textit{Bon}): Belongingness (\textit{Bel}) and Support (\textit{Sup}); 3) Activities proffered (\textit{Pro}); and 4) Clarity (\textit{Cla}): Rules (\textit{Rul}) and Values (\textit{Val}). 
        Constructs associated to \textit{PYD}: Optimism (\textit{Opt}), Pessimism (\textit{Pes}), General Self-efficacy (GSe), Agency (\textit{Age}), Comprehensibility (\textit{Com}), Manageability (\textit{Man}), and Meaningfulness (\textit{Mea}).
    }
    \label{fig:sem_figure}
\end{figure}

First, given that the focus is on discrete BNM, the scores, $S_i$, for the latent variables are generated for each individual $i$ based on the model proposed by, \citet{balaguer_2022}, as shown in Figure \ref{fig:sem_figure}. Then, these scores are discretized into quintiles, $Q_i$, following the process:

\[
Q_i = \begin{cases} 
1 & \text{if } S_i \leq P_{20} \\
2 & \text{if } P_{20} < S_i \leq P_{40} \\
3 & \text{if } P_{40} < S_i \leq P_{60} \\
4 & \text{if } P_{60} < S_i \leq P_{80} \\
5 & \text{if } S_i > P_{80}
\end{cases}
\]

where $P_{20}, P_{40}, P_{60},$ and $P_{80}$ represent the 20th, 40th, 60th, and 80th percentiles of the score distribution $S_i$, respectively.

With the discretized values, under the structure given by the \textit{PYD} model defined by the original authors, the model was re-estimated using two algorithms for BNM: the Expectation Maximization (EM) algorithm, \citet{pearl_1988} and the Bayesian Dirichlet equivalent uniform (BDeu) score, \citet{heckerman_1995}.

EM focuses on maximizing the likelihood or posterior probability of parameters given data. Designed for incomplete data, it is also effective with complete data, particularly for efficiently estimating parameters in complex models. BDeu, using a uniform prior and focusing on the BDE score, is specifically designed for model selection and parameter estimation in BNM. This approach aligns with incorporating prior knowledge explicitly and facilitates comparing different network structures within the Bayesian framework. Thus, using BDeu is expected to perform better in estimation by significantly weighting prior values, especially when using SEM analysis results as initial values.

Once the estimates were obtained, the process of generating inferences began. In BNM the inference process leverages the network's structure and the estimated conditional probability (CP) distributions. Thus, the performance of the models was evaluated using classic indicators such as Accuracy, Recall, and F1-Score, for complete data (without missing data), both in the training set (70\%, $n=710$) and the validation set (30\%, $n=305$).

The selected model was evaluated using information gain (entropy) measurements. This statistic allows understanding the reduction in uncertainty about one variable given the knowledge of another. This is quantified by comparing a variable's entropy with its conditional entropy given another variable.

The entropy of a variable measures the uncertainty or randomness associated with the variable. It is calculated using:
\begin{equation}
    H(X) = -\sum_{x \in X} P(x) \log P(x),
\end{equation}
where \(P(x)\) is the probability of each state of \(X\).

Conditional entropy measures the amount of uncertainty remaining in a variable when the state of another variable is known. It is defined as:
\begin{equation}
    H(X|Y) = -\sum_{y \in Y} P(y) \sum_{x \in X} P(x|y) \log P(x|y),
\end{equation}
where \(P(x|y)\) is the conditional probability of \(X\) given \(Y\).

Lastly, Information Gain (IG) quantifies the reduction in uncertainty about variable \(X\) due to the knowledge of variable \(Y\) and is calculated as the difference between the entropy of \(X\) and the conditional entropy of \(X\) given \(Y\):
\begin{equation}
    IG(X;Y) = H(X) - H(X|Y).
\end{equation}
A higher value of \(IG(X;Y)\) indicates that knowing \(Y\) significantly reduces the uncertainty in \(X\).

This analysis became pivotal in predicting \textit{PYD}, with a focus on the examination of conditional probabilities and the information gain of variables related to it. The visualization of BNM and the generation of CP tables were conducted to provide insights into the network's structure and the relationships between variables.

\section{Theoretical model}\label{theo_mod}

\subsection*{Personal Positive Youth Development (\textit{PYD})}
Positive psychology, especially the \textit{PYD} approach, focuses on human development through competence improvement. Influencing various research areas like education, it complements prevention efforts against instability, conflicts, or risk behaviors, promoting positive human development aspects. Both prevention and promotion perspectives are necessary for the health and wellbeing of youths, \citet{conway_2015, orejudo_2013, oliva_2011, gutierrez_2013}. The Five Cs model by Lerner—Competence, Confidence, Connection, Character, and Caring—is one of the most comprehensive, incorporating social, moral, and personal dimensions. It underscores the development of cognitive, behavioral, and social competences, with empirical support for its role in youth flourishing and positive community contribution, \citet{benson_2006, lerner_2005, lerner_2009, oliva_2010, dvorsky_2018, oliva_2011}. Based on the Five Cs, the \textit{PYD} approach centers on personal competencies crucial for youth development. It emphasizes adaptive responses to developmental tasks, with personal beliefs, skills, and capabilities as the core. This focus aims to enhance the understanding of individual factors related to wellness and adolescent context adjustment, \citet{balaguer_2020b, oliva_2010, orejudo_2013, orejudo_2021, delafuente_2017}.

\section*{Structural Model for \textit{PYD}}
The development of adolescents is significantly influenced by their immediate socialization contexts, notably family and school. Parenting practices integrating affect and communication and appropriate control support child development. Similarly, schools contribute to \textit{PYD} by fostering a supportive climate, clear rules, and positive relationships with adults, enhancing students' competency development and adjustment, \citet{white_2000, vansteenkiste_2013, baumrind_1967, maccoby_1983, barber_2005, oliva_2007, darling_1999, moos_1987, trianes_2006, marjoribanks_1980, greenberg_2003, pertegal_2015}.

Thus, the structural model aimed to examine if \textit{PYD} can be explained by the contextual variables: family and school. It was hypothesized that \textit{PYD} would be fostered by \textit{PP}—characterized by closeness and communication, promotion of autonomy, humor, and self-disclosure—and by a positive \textit{CFS}—characterized by beneficial peer relationships, a sense of belonging and support, a clear understanding of rules and values, and enriching activities and resources in the school. The empirical results show these associations and predictions among Spanish adolescents, regardless of sex and age. \textit{PP} emerges as the most significant factor in promoting \textit{PYD}, while the role of the school context is less pronounced, aligning with previous research on asset promotion in adolescence, \citet{balaguer_2020a, balaguer_2020b, oliva_2011}. These factors are interrelated, and all are associated with individual personal competencies, \citet{lerner_2005}. This approach moves beyond the individualistic view of personal competencies by including the effect of contextual variables, aligning with an interactive view of personal development and educational processes, \citet{delafuente_2017}. Family and school environments are presented as essential regulatory contexts to promote personal development. The family, as a contextual factor, is emphasized over school variables, consistent with the findings that the perception of assets is a predictor of better competence development among students, \citet{scales_2000}. Although the family context is the main factor influencing \textit{PYD}, there is also a significant association between school and \textit{PYD}, underscoring the importance of a combined effect of family and school environments in the direction of positive development.

\section{Result}\label{res}

The initial SEM showed adequate fit levels under the assumptions of, \citet{hu_1999}: comparative fit index (CFI) $>$ 0.95 and standardized root mean square residual (SRMR) $<$ 0.09; or root mean square error of approximation (RMSEA) $<$ 0.05 and SRMR $<$ 0.06, see Table \ref{table:model_fit_metrics}.

\begin{figure}[!ht]
\centering
\begin{subfigure}[b]{0.5\textwidth}
\includegraphics[width=\textwidth]{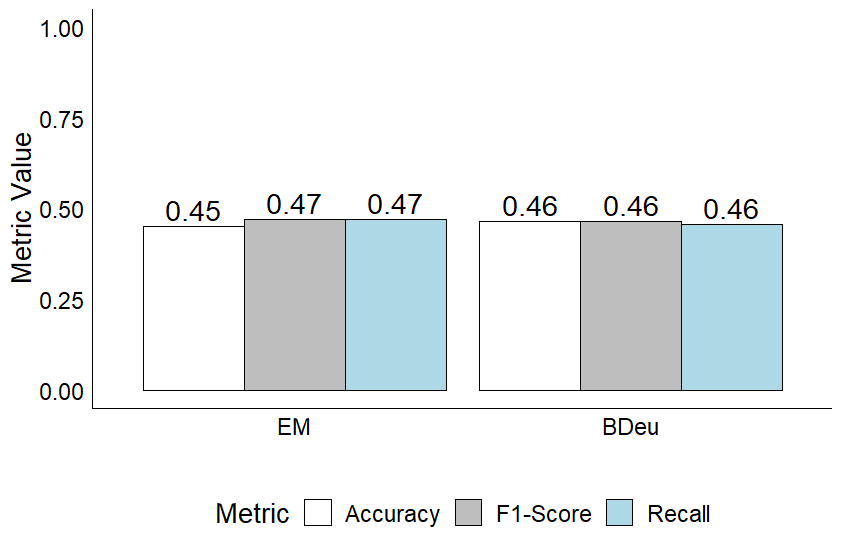}
\caption{Training Data Metrics}
\label{fig:training-metrics}
\end{subfigure}
\hfill
\begin{subfigure}[b]{0.5\textwidth}
\includegraphics[width=\textwidth]{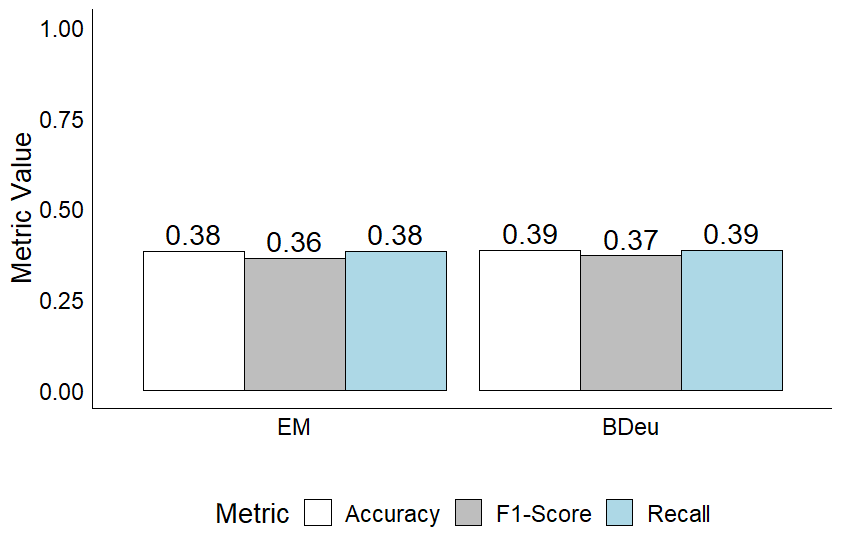}
\caption{Validation Data Metrics}
\label{fig:validation-metrics}
\end{subfigure}
\caption{Comparative Performance Metrics for Bayesian Network Fitting Methods}
\label{fig:both-metrics}
\end{figure}

In this manner, the scores for each of the latent variables from the SEM were obtained, see Table \ref{table:factor_loadings_summary}. Subsequently, using these scores and their translation into discrete values represented by the quintiles of their distribution, the conditional probability tables were generated, and from these, the inferences that were utilized to estimate the fit statistics of the BNM for each evaluated algorithm, Figure \ref{fig:both-metrics}. From this analysis, it is observed that generally, the algorithms provide similar results; however, the BDeu method, as expected given its emphasis on initial values, demonstrated better performance than the EM, which is why it was chosen to perform the final inferences on the entire data table. It is clear that the fit values are very low; this is due to the model being required to handle high complexity by dividing responses into quintiles. However, the interest before adjustment is to evaluate the model's ability to convert factor scores into ordinal variables that are interpretable from a theoretical perspective. Nevertheless, the model was also evaluated with the minimum number of levels (2), which corresponds to the division given by the median. In that case, the values of the three metrics reached 80\%.

\begin{figure}[!ht]
\centering
\includegraphics[trim={1mm 0 0 0},clip,width=\linewidth]{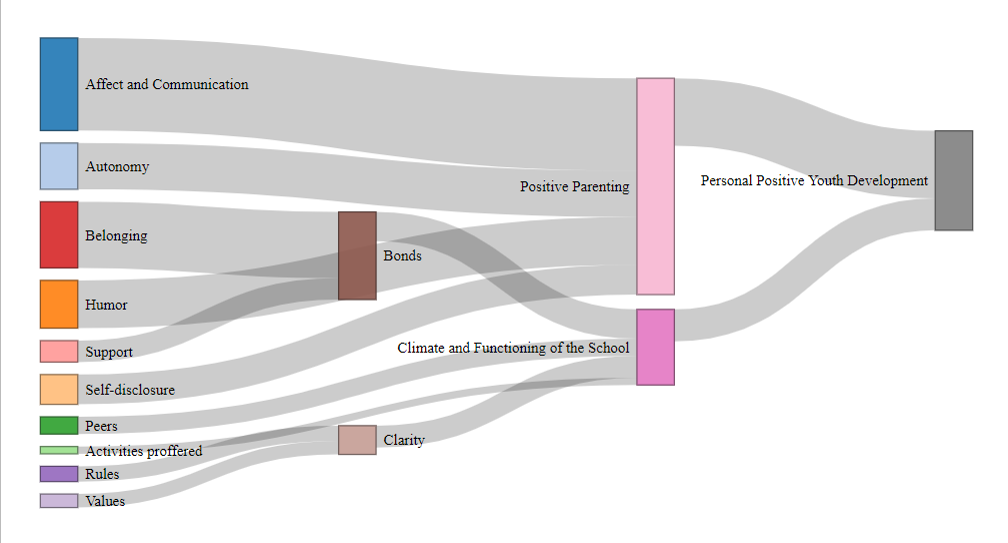}
\caption{Diagram Representing Information Gain from Latent Variables to Personal Positive Youth Development (\textit{PYD}).}
\label{fig:sankey}
\end{figure}

The information gain values for the latent variables of the model are presented in Figure \ref{fig:sankey}, where the effect of positive parenting is highlighted, within which \textit{AfC} is the construct or latent variable contributing the most. Regarding the \textit{CFS}, the \textit{Bon}, represented to a greater extent by belonging, show the greatest contribution to the target variable, \textit{PYD}.

\begin{figure}[!ht]
\centering
\begin{subfigure}{.3\textwidth}
  \centering
  \includegraphics[width=\linewidth]{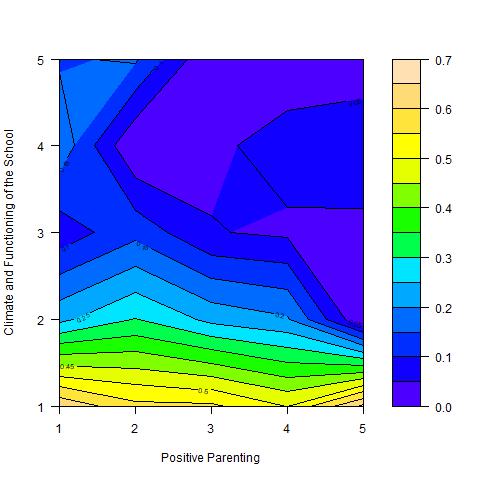}
  \caption{\textit{PYD} Level 1}
  \label{fig:sub1}
\end{subfigure}%
\begin{subfigure}{.3\textwidth}
  \centering
  \includegraphics[width=\linewidth]{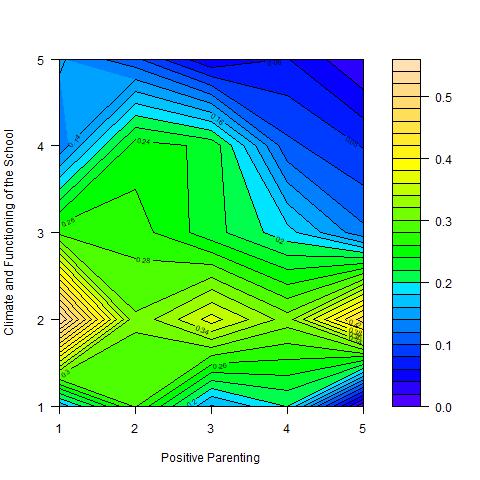}
  \caption{\textit{PYD} Level 2}
  \label{fig:sub2}
\end{subfigure}
\begin{subfigure}{.3\textwidth}
  \centering
  \includegraphics[width=\linewidth]{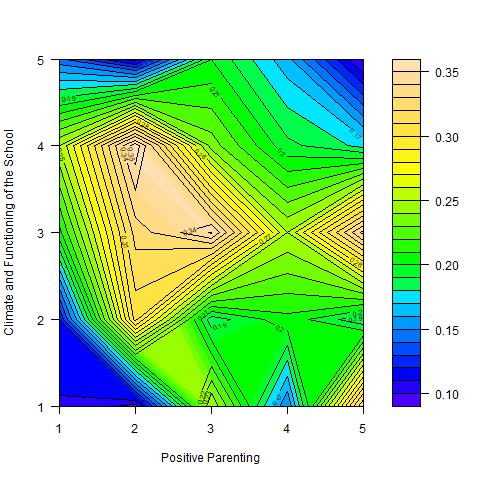}
  \caption{\textit{PYD} Level 3}
  \label{fig:sub3}
\end{subfigure}
\begin{subfigure}{.3\textwidth}
  \centering
  \includegraphics[width=\linewidth]{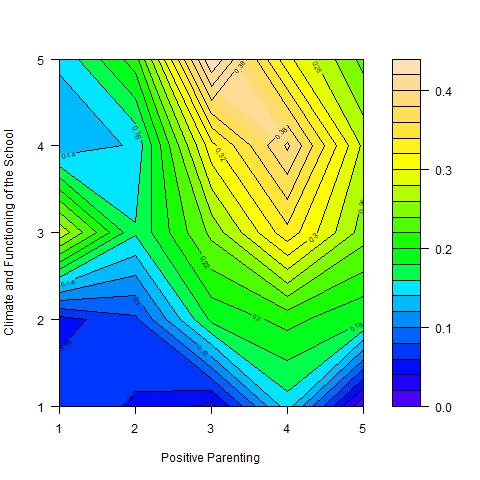}
  \caption{\textit{PYD} Level 4}
  \label{fig:sub4}
\end{subfigure}
\begin{subfigure}{.3\textwidth}
  \centering
  \includegraphics[width=\linewidth]{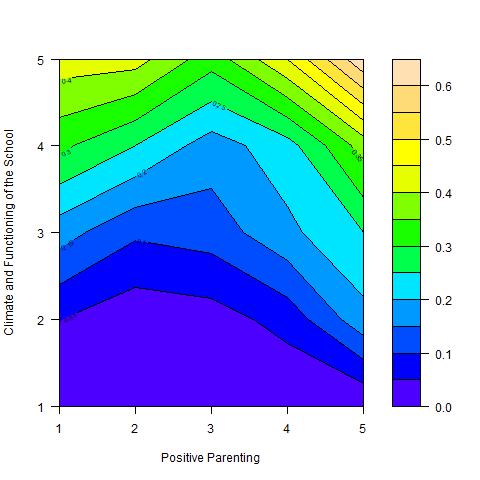}
  \caption{\textit{PYD} Level 5}
  \label{fig:sub5}
\end{subfigure}
\caption{Probability Contour Plots for Each Level of Personal Positive Youth Development (\textit{PYD}) showing its relationship with Positive Parenting (\textit{PP}) and Climate and Functioning of the School (\textit{CFS}).}
\label{fig:test}
\end{figure}

Lastly, to evaluate the nonlinear behaviors of the assessed factors, contour figures were generated, as seen in Figure \ref{fig:test}. These figures display the conditional probability tables for each level (quintile) of the latent variables \textit{PYD}, based on the quintiles of the \textit{PP} and \textit{CFS}.

For the first level of \textit{PYD}, it is observed that its greatest conditional dependency comes from the lower values of \textit{CFS}, a situation not observed with \textit{PP}. This finding, not evident in the SEM model, indicates that the linear relationships implied by the SEM model—that low levels of \textit{PP} and \textit{CFS} would conditionally lead to low \textit{PYD} levels—may not hold. This suggests a reevaluation of assuming parenting styles have a linear effect.

At the second level of \textit{PYD}, a similar pattern is observed, though \textit{PP} starts to show a slight effect on \textit{PYD}. However, it is clear that the probability of being at level 2 of \textit{PYD} almost exclusively depends on \textit{CFS}.

For the third level of \textit{PYD}, \textit{CFS} begins to interact with \textit{PP}, where the highest probability of being at \textit{PYD} level three depends on both levels two and three of \textit{PP}, as well as levels 2, 3, and 4 of \textit{CFS}.

At the fourth level of \textit{PYD}, the proportional effect of \textit{PP} and \textit{CFS} consolidates, though an effect less pronounced at level 3 becomes evident. At this level, changes in \textit{CFS} show less steep slopes than those in \textit{PP}, meaning improvements in \textit{PP} level have a more pronounced effect on the probability of \textit{PYD} level 4 than changes in \textit{CFS} levels do.

Finally, the fifth level of \textit{PYD} somewhat replicates the behavior of level 1, where the most significant changes are due to increases in \textit{CFS} levels. However, unlike \textit{PYD} 1, this level shows that the highest probability values are achieved at high levels of both \textit{PP} and \textit{CFS}. This somewhat aligns with the SEM results, where both factors additively and positively influence \textit{PYD}.

Overall, the effect of \textit{CFS} is somewhat proportional to the levels of \textit{PYD}; contrary to the effect of \textit{PP}, which at its lower levels does not significantly impact the target variable. Its most significant impact is observed starting from the third level of \textit{PYD}, reaching its maximum expression at the highest level of \textit{PYD}.

\begin{figure}[!ht]
  \centering
  \begin{subfigure}{0.45\textwidth}
    \includegraphics[width=\linewidth, clip]{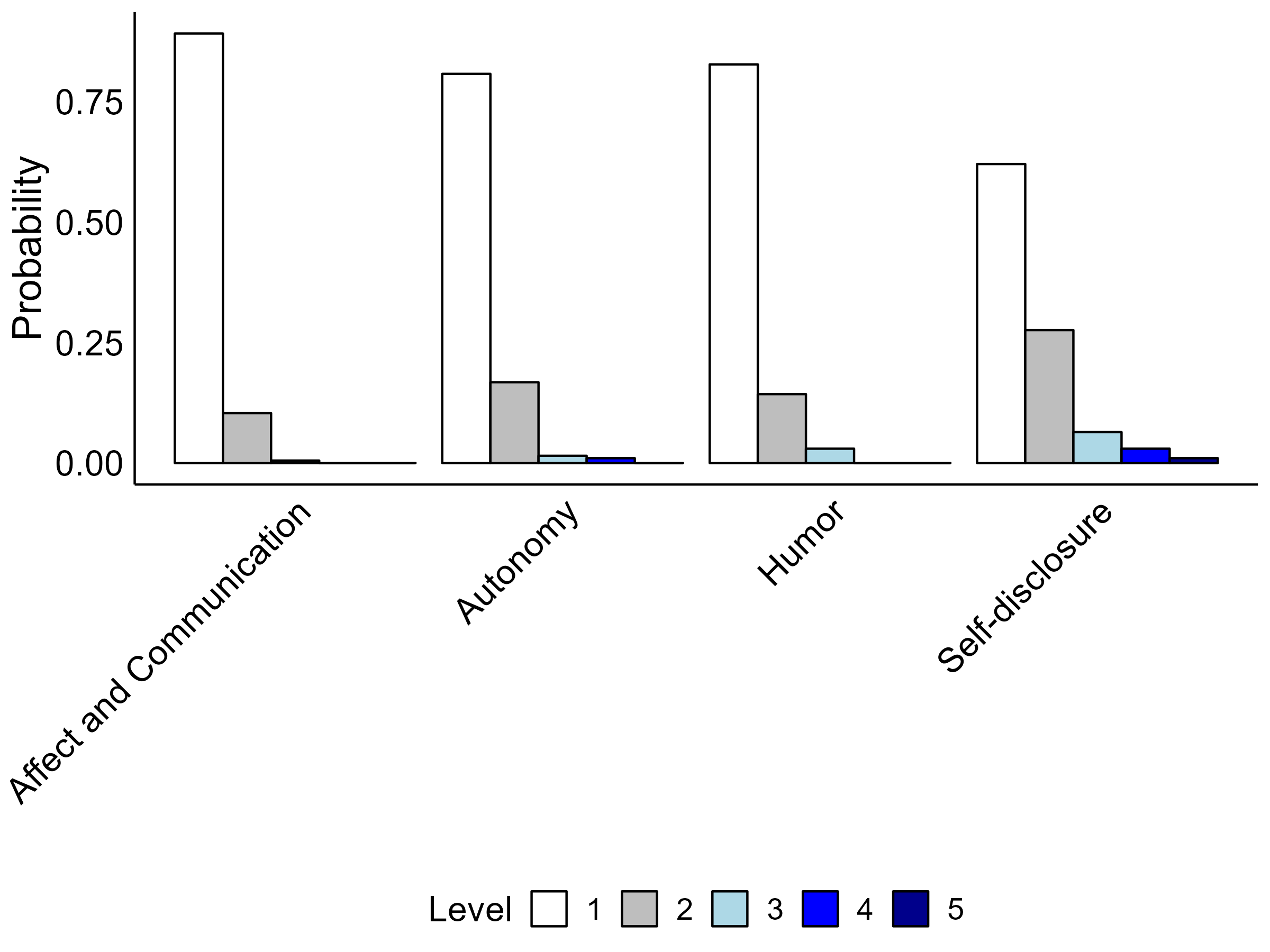}
    \caption{Level 1}
    \label{subfig:pp_level_1}
  \end{subfigure}
  \hspace{-5mm}
  \begin{subfigure}{0.45\textwidth}
    \includegraphics[width=\linewidth, clip]{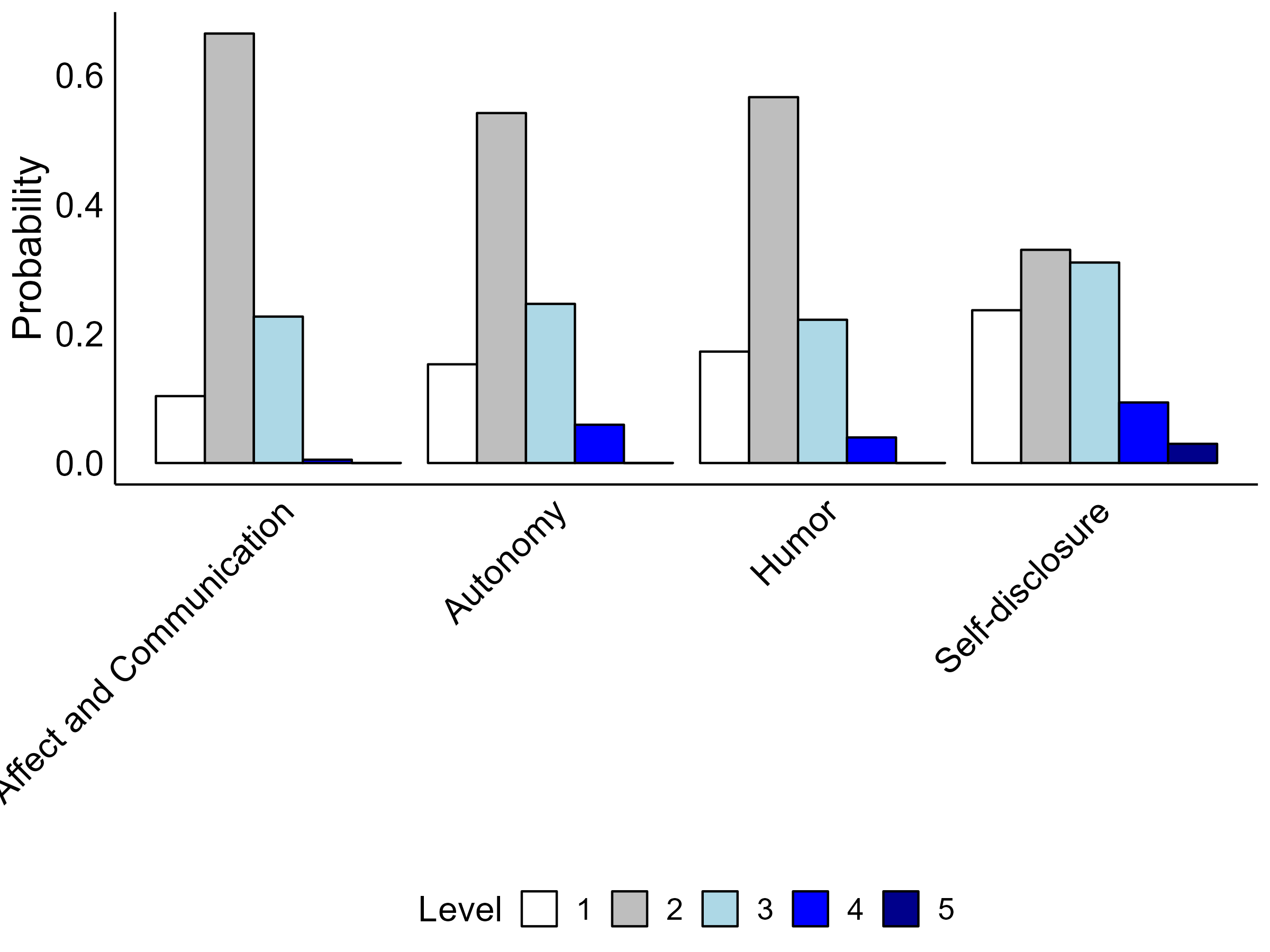}
    \caption{Level 2}
    \label{subfig:pp_level_2}
  \end{subfigure}
  \vspace{5mm}
  \begin{subfigure}{0.45\textwidth}
    \includegraphics[width=\linewidth, clip]{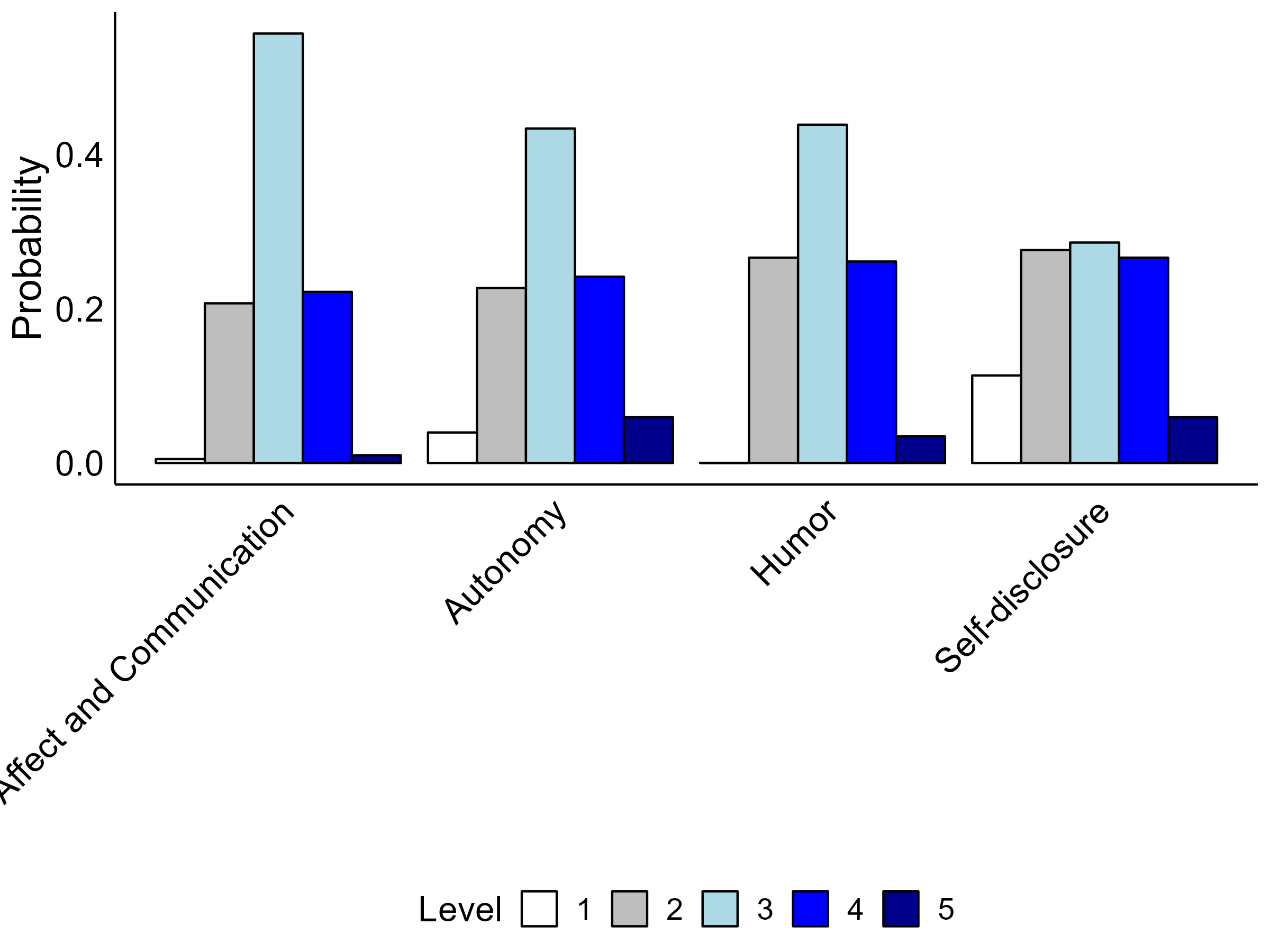}
    \caption{Level 3}
    \label{subfig:pp_level_3}
  \end{subfigure}
  \hspace{-5mm}%
  \begin{subfigure}{0.45\textwidth}
    \includegraphics[width=\linewidth, clip]{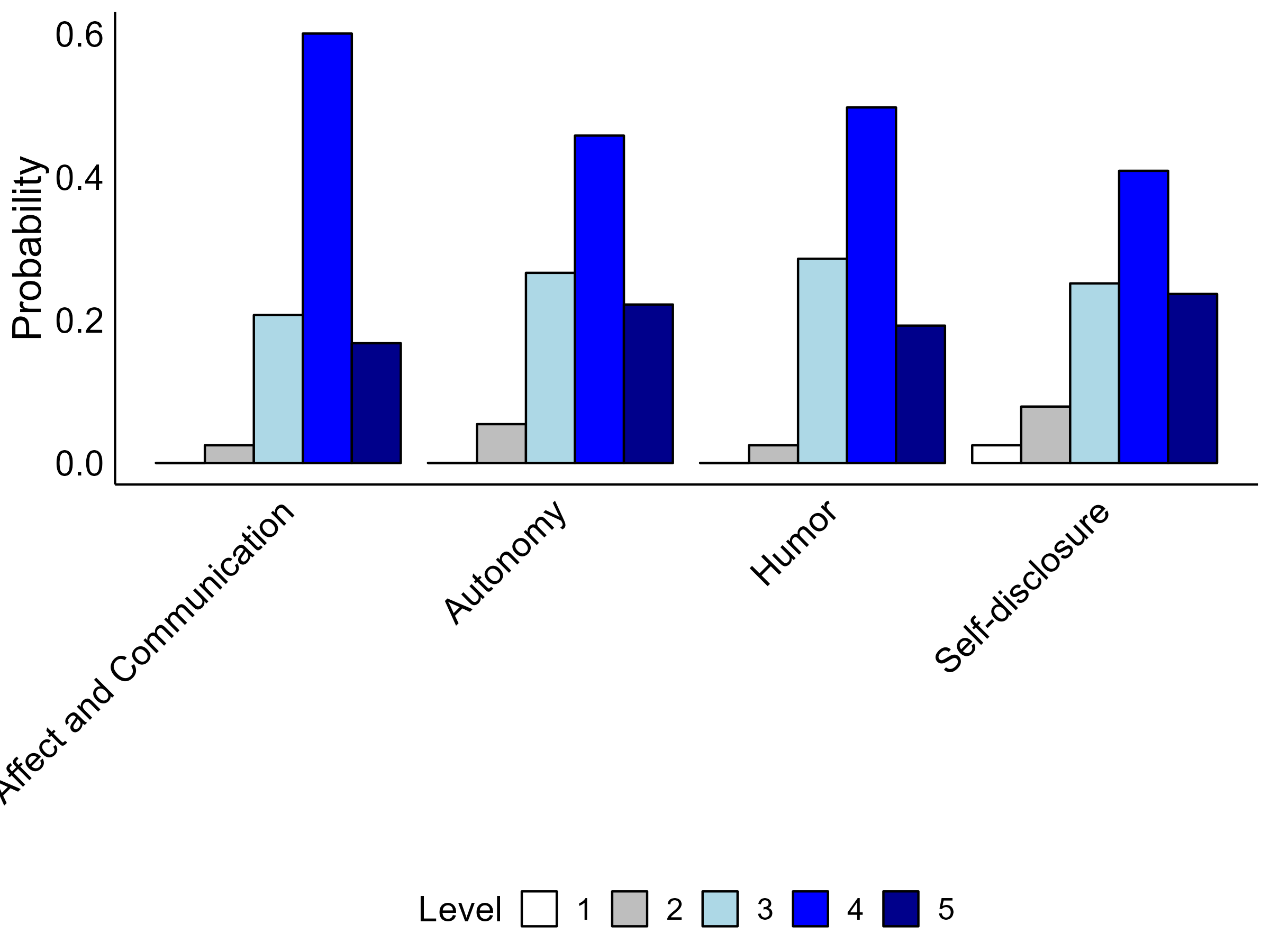}
    \caption{Level 4}
    \label{subfig:pp_level_4}
  \end{subfigure}
  
  \vspace{5mm}
  \begin{subfigure}{0.45\textwidth}
    \includegraphics[width=\linewidth, clip]{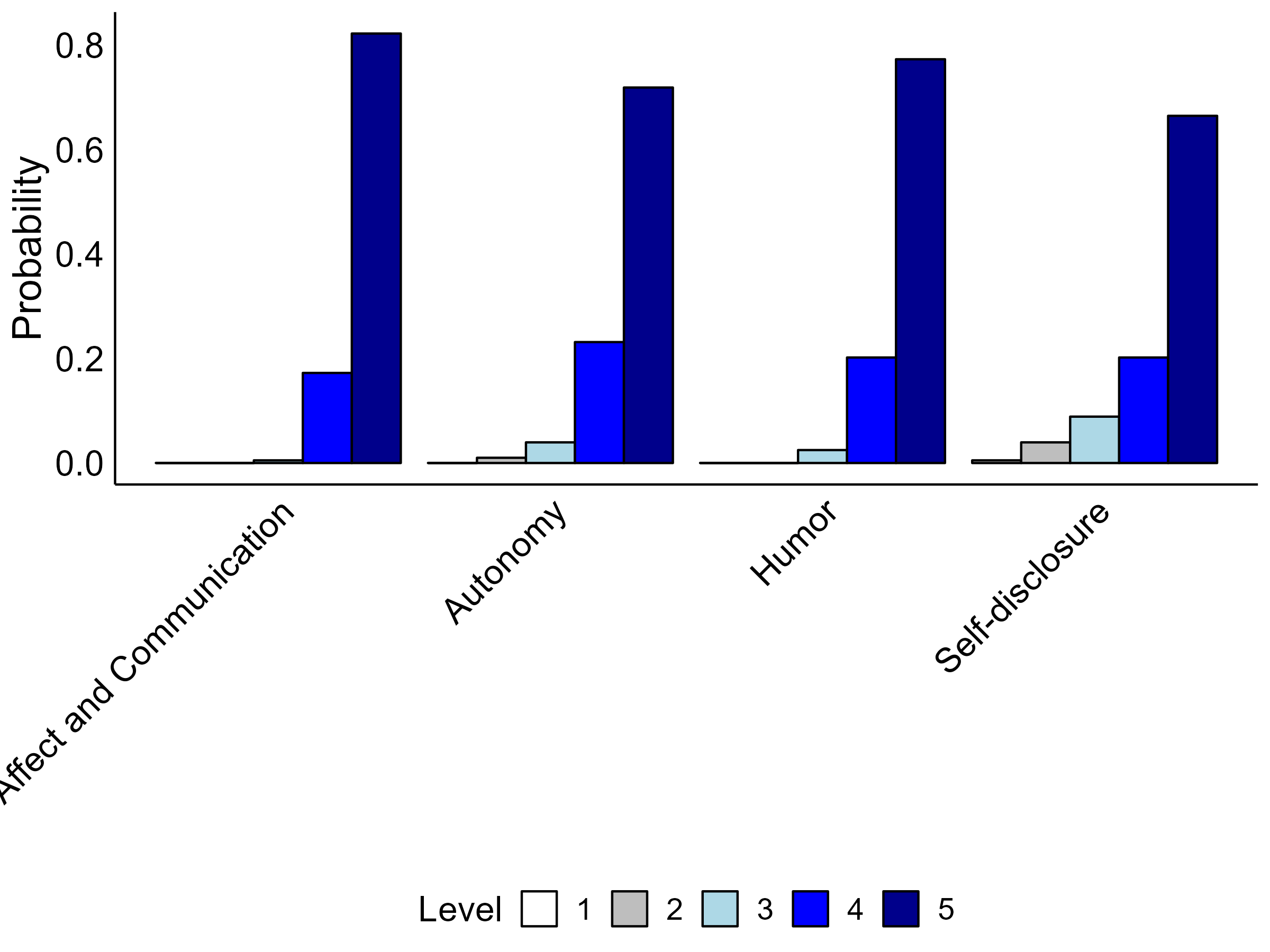}
    \caption{Level 5}
    \label{subfig:pp_level_5}
  \end{subfigure}
  \caption{Conditional Probability Distributions of Variables Influencing \textit{PP}}
  \label{fig:pp_distributions}
\end{figure}

\begin{figure}[!ht]
  \centering
  \begin{subfigure}{0.45\textwidth}
    \includegraphics[width=\linewidth, clip]{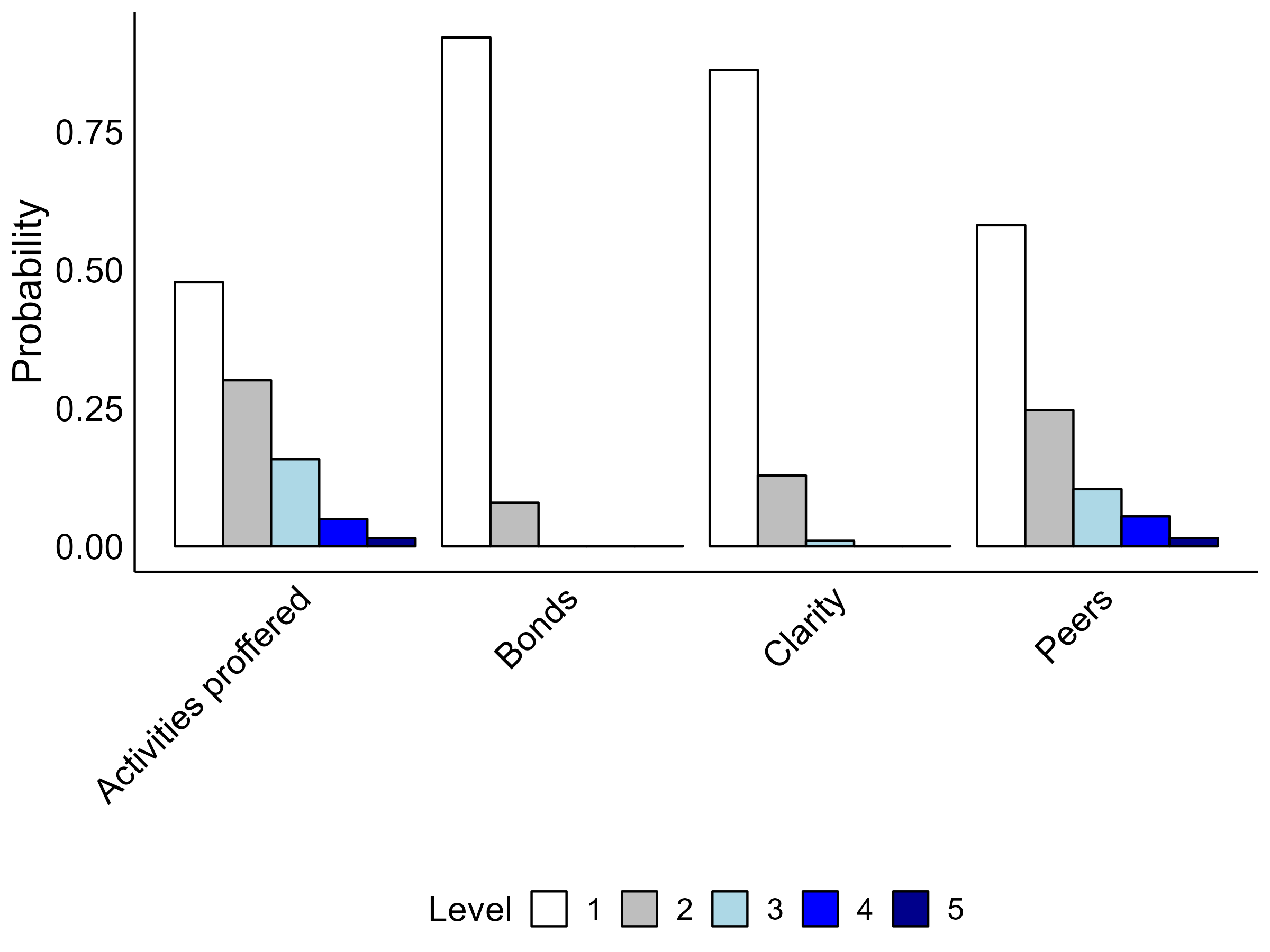}
    \caption{Level 1}
    \label{subfig:pcfs_level_1}
  \end{subfigure}\hspace{-5mm}%
  \begin{subfigure}{0.45\textwidth}
    \includegraphics[width=\linewidth, clip]{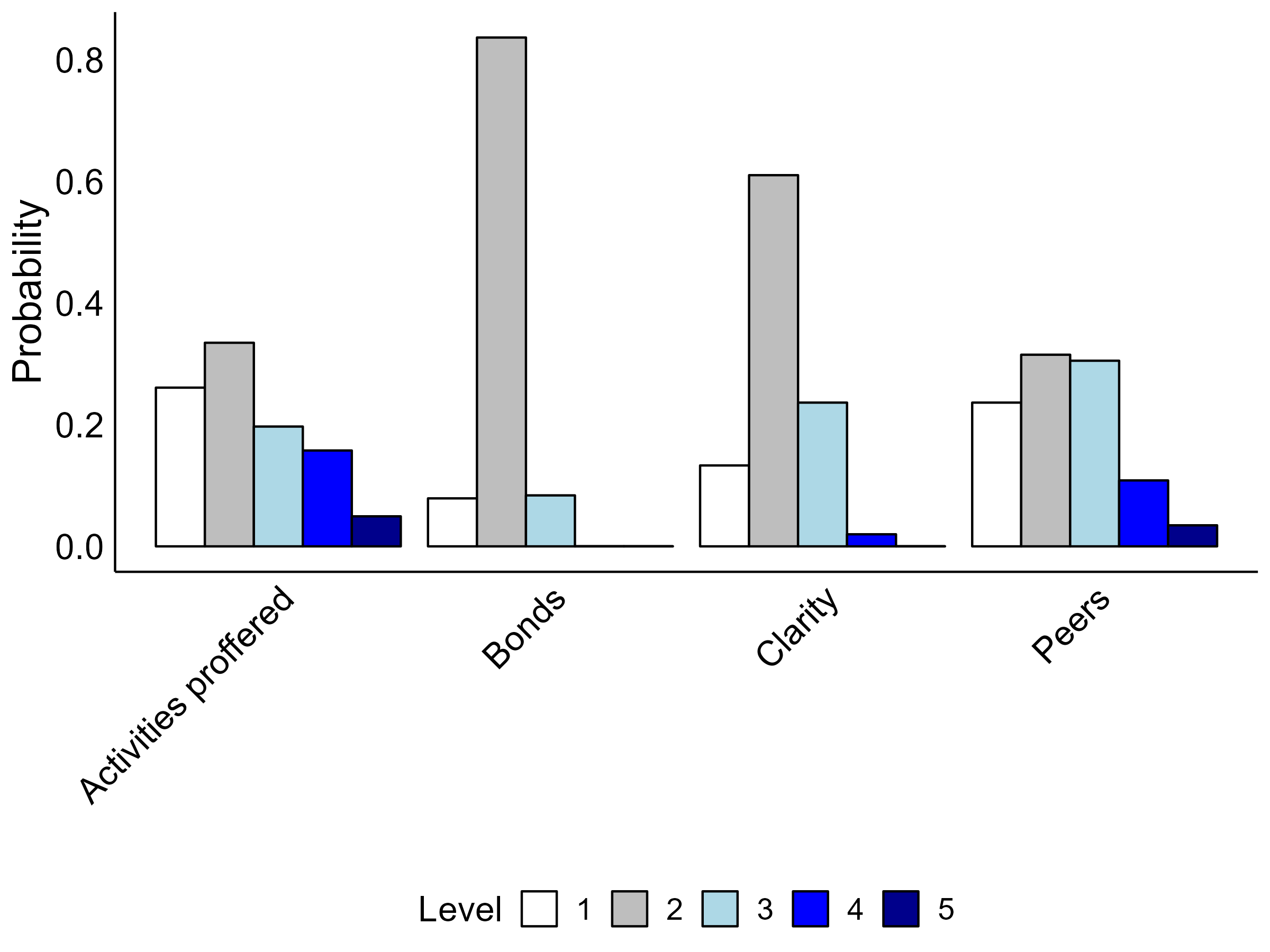}
    \caption{Level 2}
    \label{subfig:pcfs_level_2}
  \end{subfigure}
  
  \vspace{5mm}
  \begin{subfigure}{0.45\textwidth}
    \includegraphics[width=\linewidth, clip]{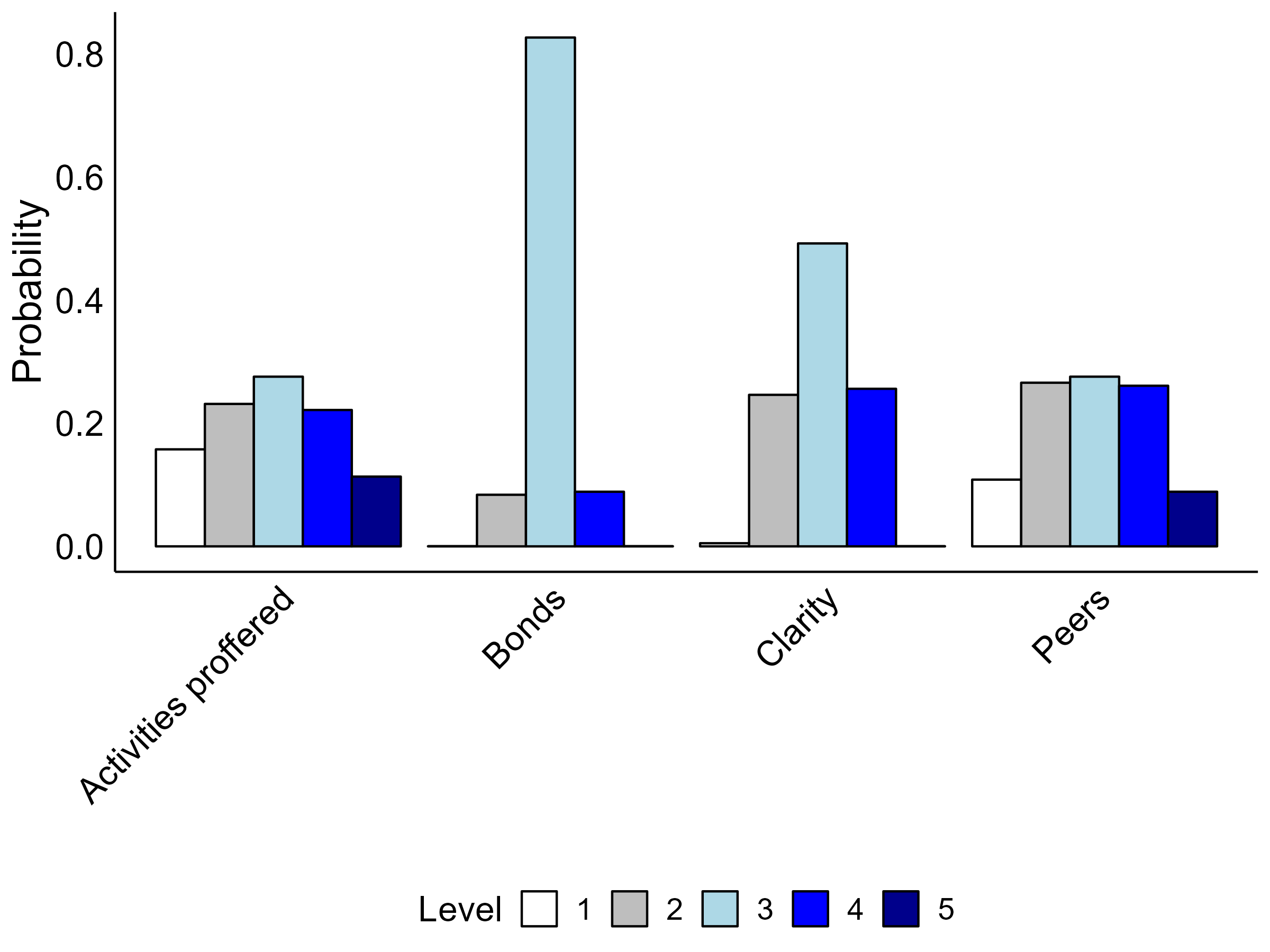}
    \caption{Level 3}
    \label{subfig:pcfs_level_3}
  \end{subfigure}\hspace{-5mm}%
  \begin{subfigure}{0.45\textwidth}
    \includegraphics[width=\linewidth, clip]{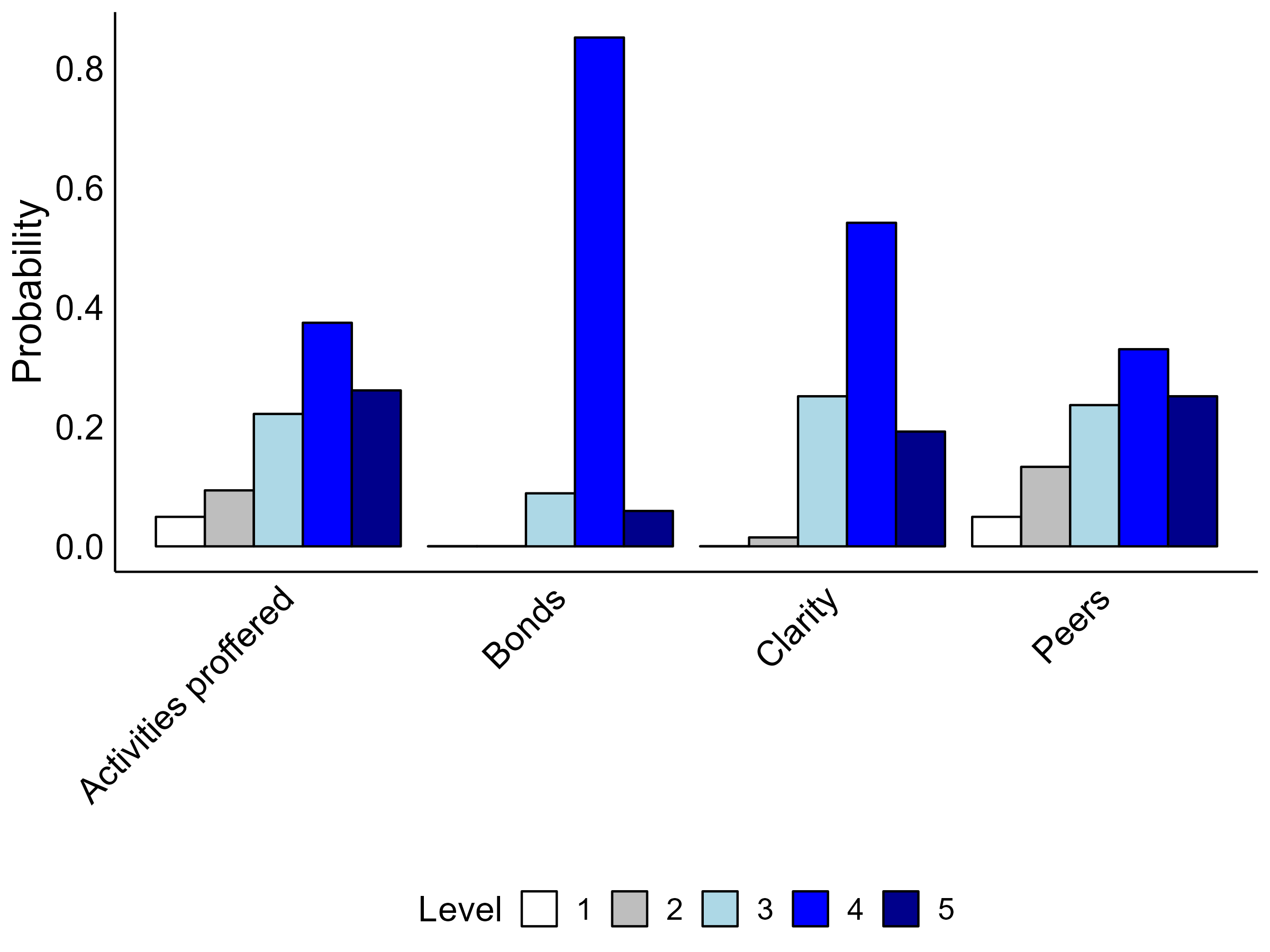}
    \caption{Level 4}
    \label{subfig:pcfs_level_4}
  \end{subfigure}
  
  \vspace{5mm}
  \begin{subfigure}{0.45\textwidth}
    \includegraphics[width=\linewidth, clip]{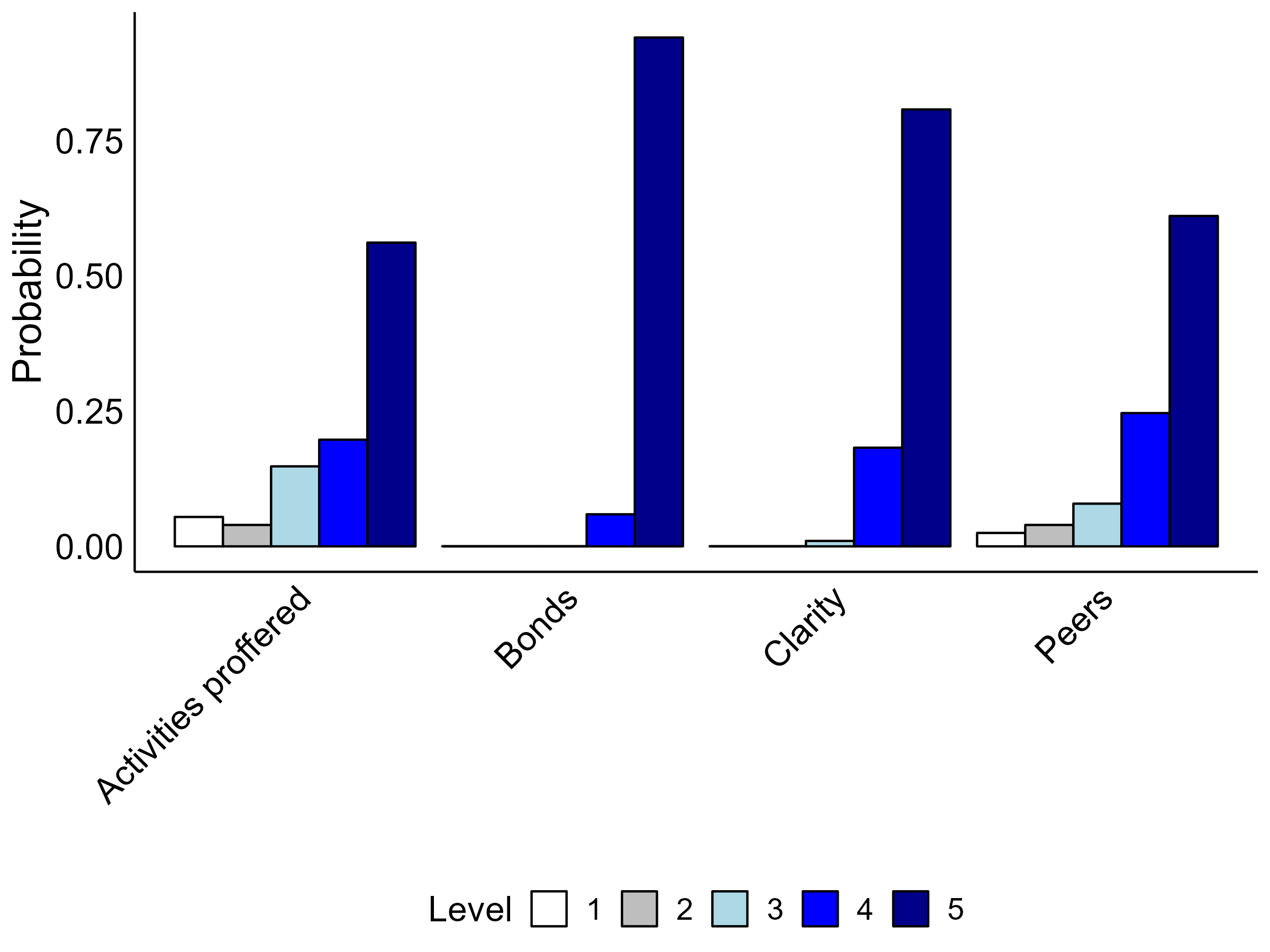}
    \caption{Level 5}
    \label{subfig:pcfs_level_5}
  \end{subfigure}
  \caption{Conditional Probability Distributions of Variables Influencing \textit{CFS}}
  \label{fig:pcfs_distributions}
\end{figure}

\begin{figure}[!ht]
  \centering
  \begin{subfigure}{0.45\textwidth}
    \includegraphics[width=\linewidth, clip]{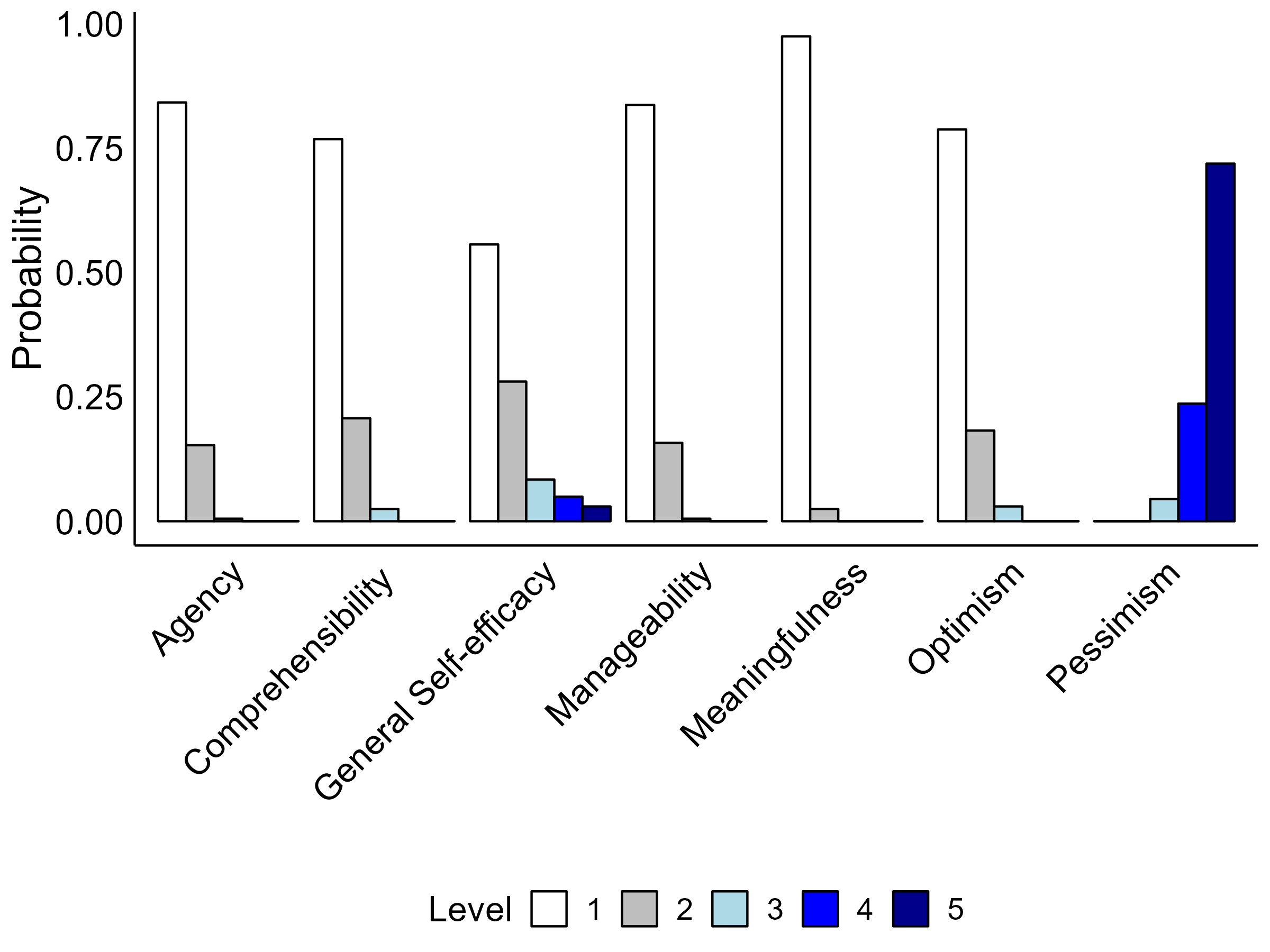}
    \caption{Level 1}
    \label{subfig:ppyd_level_1}
  \end{subfigure}\hspace{-1mm}%
  \begin{subfigure}{0.45\textwidth}
    \includegraphics[width=\linewidth, clip]{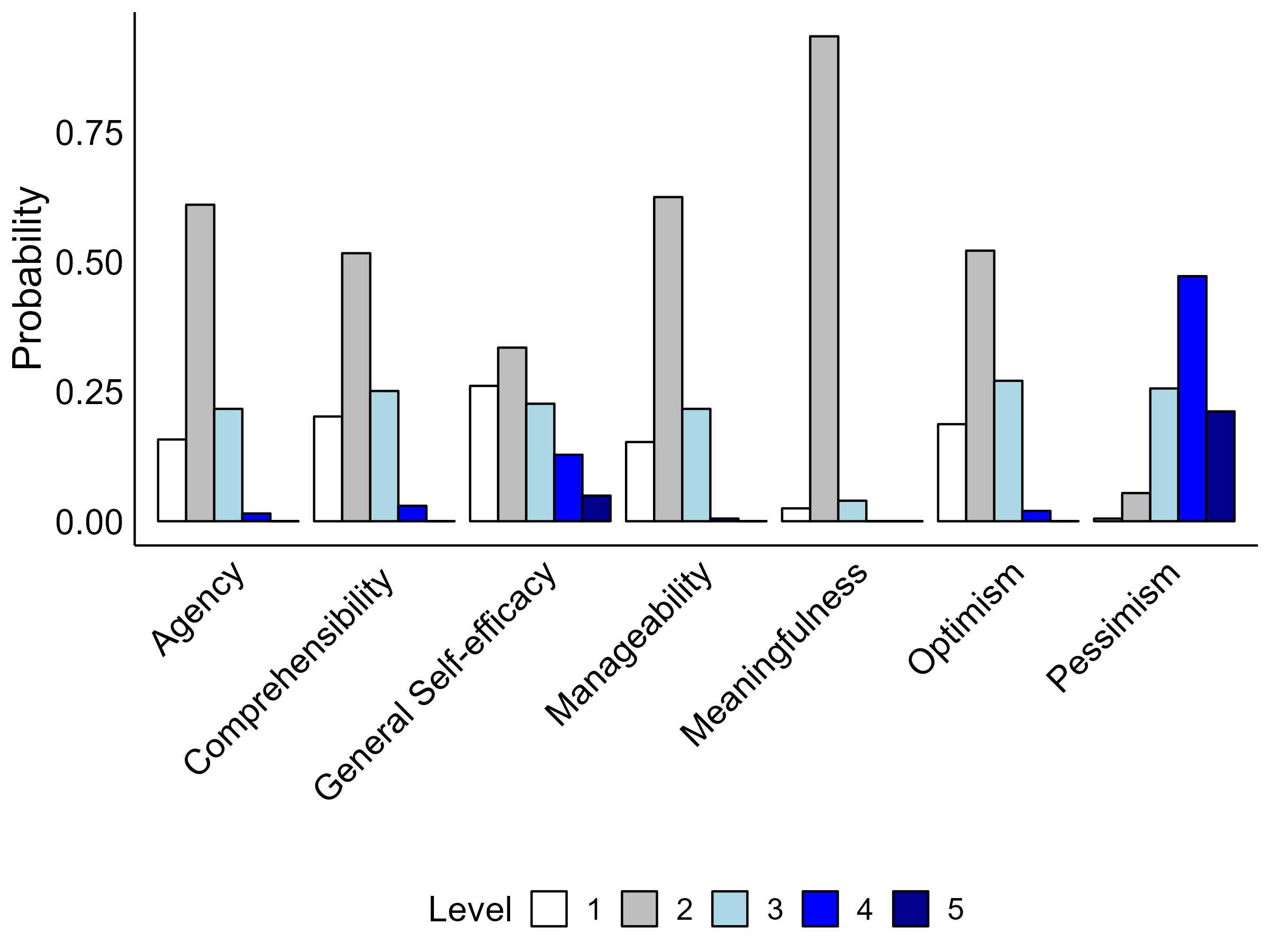}
    \caption{Level 2}
    \label{subfig:ppyd_level_2}
  \end{subfigure}
  
  \vspace{5mm}
  \begin{subfigure}{0.45\textwidth}
    \includegraphics[width=\linewidth, clip]{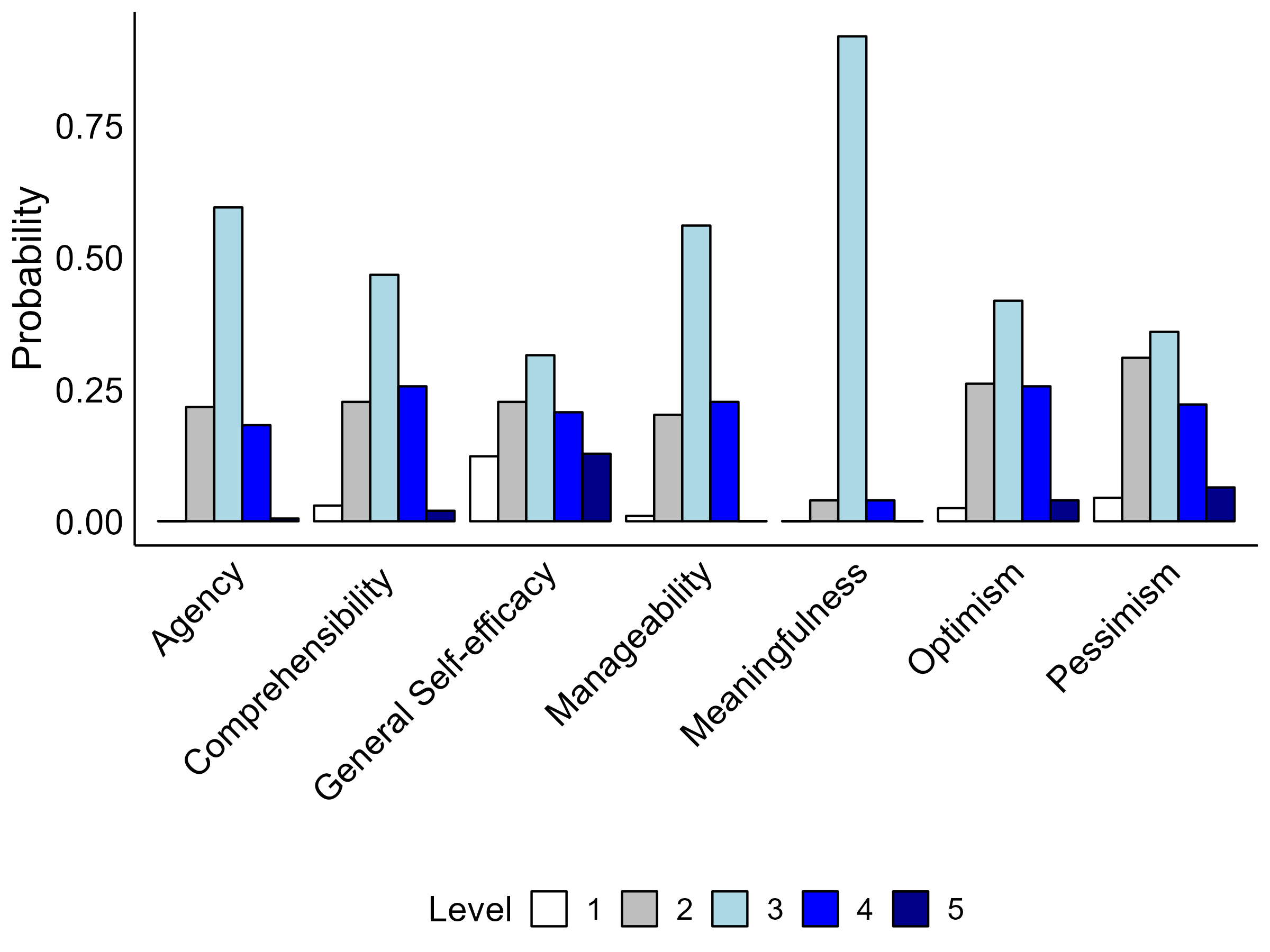}
    \caption{Level 3}
    \label{subfig:ppyd_level_3}
  \end{subfigure}\hspace{-1mm}%
  \begin{subfigure}{0.45\textwidth}
    \includegraphics[width=\linewidth, clip]{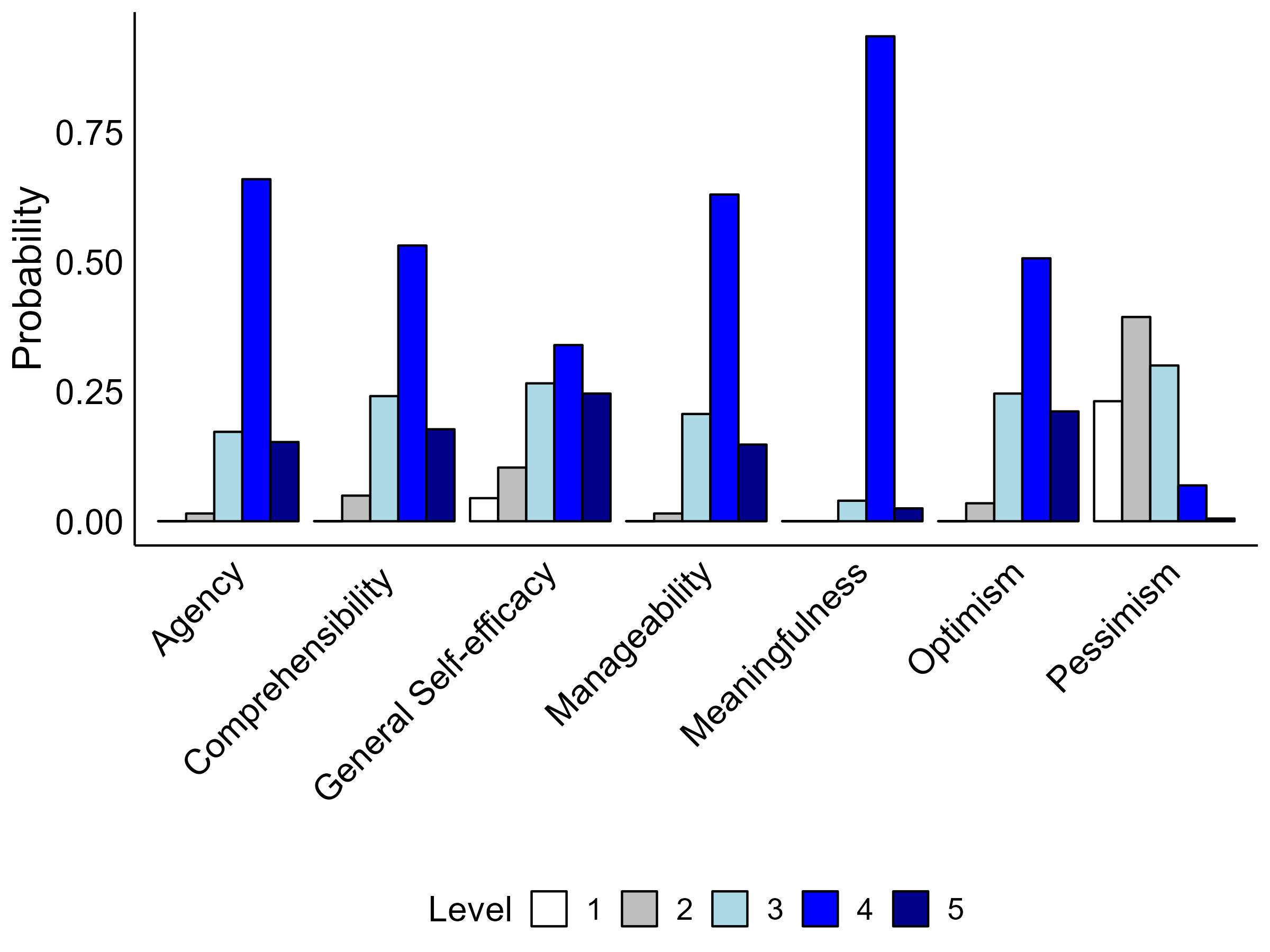}
    \caption{Level 4}
    \label{subfig:ppyd_level_4}
  \end{subfigure}
  
  \vspace{5mm}
  \begin{subfigure}{0.45\textwidth}
    \includegraphics[width=\linewidth, clip]{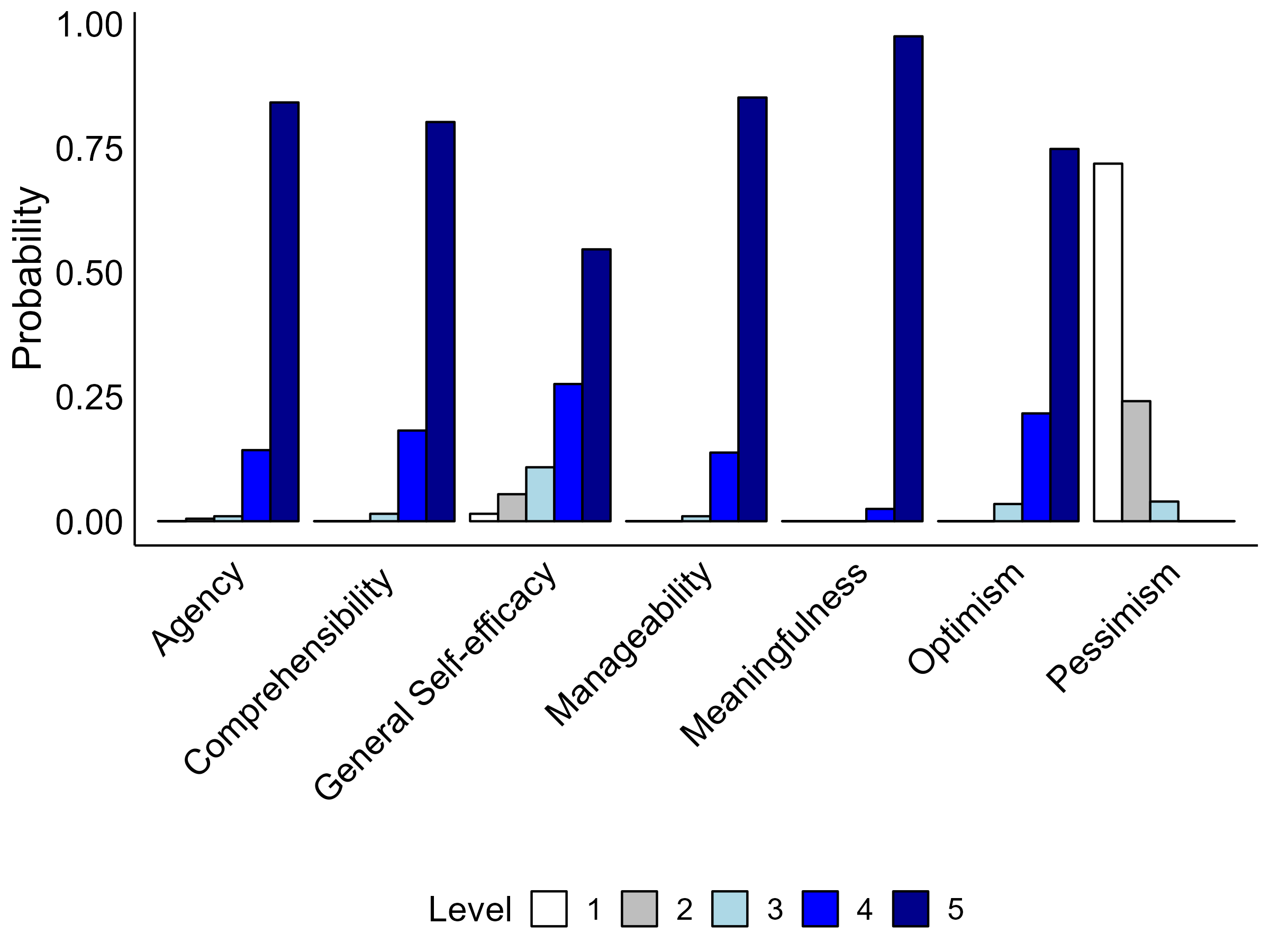}
    \caption{Level 5}
    \label{subfig:ppyd_level_5}
  \end{subfigure}
  \caption{Conditional Probability Distributions of Variables Influencing \textit{PYD}}
  \label{fig:ppyd_distributions}
\end{figure}

The possibilities for diagnosis within a network of this style are numerous and depend on the researcher's interest in which explicit effects they want to evaluate on specific target variables. For this case, and as an example of the potential of the analysis of conditional probabilities, the relationship between the second-level latent variables in relation to the target variables \textit{PP}, \textit{CFS}, and \textit{PYD} was analyzed, see Figures \ref{fig:pp_distributions}, \ref{fig:pcfs_distributions}, and \ref{fig:ppyd_distributions}.

For the second-level variables \textit{Aut}, \textit{Hum}, \textit{Dis}, and \textit{AfC} related to \textit{PP}, as shown in Figure \ref{fig:pp_distributions}, it is noted that these variables generally exhibit the same behavior regardless of the level of \textit{PP} assessed; overall, \textit{AfC} shows the highest conditional probabilities and \textit{Dis} the lowest. It is also highlighted that the outcome is as expected, and it replicates the linear effect predicted by the SEM, that is, each level of \textit{PP} is directly related to the levels of the variables it influences, only \textit{Dis} does not clearly show this relationship at levels two and three, as seen in Figures \ref{subfig:pp_level_2} and \ref{subfig:pp_level_3}, with levels two and three of \textit{Dis} having the same impact on levels two and three of \textit{PP}. In other words, levels two and three of \textit{Dis} do not discriminate in their ability to identify intermediate levels of \textit{PP}.

For \textit{CFS}, the variables with the best predictive capacity are, in order, \textit{Bon} and \textit{Cla}, Figure \ref{fig:pcfs_distributions}, and similarly to \textit{PP}, here the variables with more problems to discriminate intermediate values of \textit{CFS} are \textit{Pro} and \textit{Pee}, Figures \ref{subfig:pcfs_level_2}, \ref{subfig:pcfs_level_3} and \ref{subfig:pcfs_level_4}.

Regarding \textit{PYD} and its second-level variables, the behavior does not present as much indefiniteness as with the previous constructs, as shown in Figure \ref{fig:ppyd_distributions}. The predictive effect of \textit{Mea} and the low prediction of \textit{Gse} are highlighted. Moreover, \textit{Gse}, \textit{Opt}, and \textit{Pes} are the variables with the greatest difficulties in identifying intermediate levels of \textit{PYD}, as seen in Figures \ref{subfig:ppyd_level_3} and \ref{subfig:ppyd_level_4}.

This demonstrates the utility of complementing the use of SEM with BNM, by converting regression coefficient values $\beta$ into probabilities, which have a clearer interpretation, not only when evaluating the model but also if the interest is to apply it in fieldwork.

\FloatBarrier

\begin{table}[ht]
\centering
\small
\caption{Summary of Factor Loadings by Latent Variable in the SEM Model}
\label{table:factor_loadings_summary}
\begin{tabularx}{\textwidth}{lXccc}
\toprule
\textbf{Layer} & \textbf{Latent Variable} & \textbf{Mean Load} & \textbf{Min Load} & \textbf{Max Load} \\
\midrule
First & Personal Positive Youth Development (\textit{PYD}) & 0.592 & -0.789 & 0.979 \\
 & Perception of the Climate and Functioning of the School (\textit{CFS}) & 0.623 & 0.122 & 0.950 \\
 & Positive Parenting (\textit{PP}) & 0.777 & 0.549 & 0.912 \\
\addlinespace
2nd for \textit{PP} & Affect and communication (\textit{AfC}) & 0.775 & 0.716 & 0.820 \\
 & Autonomy granting (\textit{Aut}) & 0.692 & 0.475 & 0.787 \\
 & Humor (\textit{Hum}) & 0.778 & 0.714 & 0.829 \\
 & Self-disclosure (\textit{Dis}) & 0.770 & 0.676 & 0.860 \\
\addlinespace
2nd for \textit{CFS} & Peers relations (\textit{Pee}) & 0.700 & 0.631 & 0.739 \\
& Belongingness (\textit{Bel}) & 0.875 & 0.864 & 0.888 \\
& School bonds (\textit{Bon}) & 0.849 & 0.848 & 0.850 \\
& Clarity of rules and values (\textit{Cla}) & 0.875 & 0.836 & 0.915 \\
\addlinespace
3rd for \textit{CFS} & Support (\textit{Sup}) & 0.795 & 0.713 & 0.840 \\
& Rules (\textit{Rul}) & 0.763 & 0.683 & 0.822 \\
& Values (\textit{Val}) & 0.750 & 0.694 & 0.838 \\
& Activities proffered (\textit{Pro}) & 0.774 & 0.726 & 0.823 \\
\addlinespace
2nd for \textit{PYD} & Optimism (\textit{Opt}) & 0.550 & 0.481 & 0.733 \\
& Pessimism (\textit{Pes}) & 0.648 & 0.583 & 0.698 \\
& General Self-efficacy (GSe) & 0.612 & 0.503 & 0.700 \\
& Agency (\textit{Age}) & 0.523 & 0.434 & 0.606 \\
& Comprehensibility (\textit{Com}) & 0.497 & 0.429 & 0.545 \\
& Manageability (\textit{Man}) & 0.455 & 0.363 & 0.561 \\
& Meaningfulness (\textit{Mea}) & 0.448 & 0.138 & 0.586 \\
\bottomrule
\end{tabularx}
\end{table}

\begin{table}[ht]
\centering
\caption{Model Fit Metrics for the SEM Model}
\label{table:model_fit_metrics}
\begin{tabular}{lc}
\toprule
\textbf{Metric} & \textbf{Value} \\
\midrule
Root Mean Square Error of Approximation (RMSEA) & 0.034 \\
Comparative Fit Index (CFI) & 0.906 \\
Standardized Root Mean Square Residual (SRMR) & 0.045 \\
\bottomrule
\end{tabular}
\end{table}

\FloatBarrier

\FloatBarrier

\FloatBarrier

\FloatBarrier

\FloatBarrier

\FloatBarrier

\section{Discussion}\label{dis}
The definition of an inferential causality model is inherently dependent on the researcher's hypotheses, viewing statistical and computational methodologies as instrumental yet subordinate to theoretical frameworks. This stance acknowledges that while methodological tools offer valuable insights, their evaluation must align with established theoretical requirements. In addressing the integration of BNM and SEM for causal inference, this discussion navigates from overarching methodological considerations to specific criticisms and advancements.
The main challenge in employing BNM lies in their directional dependency relationships, which often do not distinguish between association and causality. Conversely, SEM's limitation is its fixed model structure post-definition, unable to adapt to new data insights. A proposed solution involves a two-step approach: initial measurement model identification via SEM, followed by exploratory structural relationship analysis through BNM. This iterative method balances theoretical rigor with data-driven flexibility, allowing for the adjustment of structural models based on emerging data patterns.
Historical critiques often misinterpret SEM's approach to causal inference, suggesting an overreliance on associations without proper causal justification. Advances in methodological frameworks, such as graphical models and \textit{do-calculus},, \citet{pearl_2014}, have clarified SEM's role in causal analysis, stressing the necessity of differentiating statistical from causal assumptions. Despite these clarifications, the application of BNM in a purely data-driven manner remains prone to establishing unfounded causal connections, emphasizing the need for theoretical grounding and careful interpretation of conditional dependencies.

Within the specific context of \textit{PYD} and its second-level variables, the predictive effect of \textit{Mea} and the limitations of \textit{Gse} in identifying intermediate levels of \textit{PYD} were examined. These findings demonstrate the analytical value of combining SEM with BNM, transitioning from regression coefficients to probabilities for clearer interpretation. However, caution is advised in interpreting BNM' autonomously determined directionality, which may conflate probabilistic dependencies with causality. It should be noted that the model of, \citet{balaguer_2021} is used as a starting point, which considers the \textit{PYD} as an adolescent who perceives themselves as not pessimistic and, at the same time, optimistic, efficient, and competent. They are capable of setting goals and achieving them, and perceive the environment in which they move as understandable, manageable, and meaningful. It is considered by the adolescent that the perception of a \textit{PP} is characterized by affection and communication (\textit{AfC}) with their parents, who also promote their autonomy, possess good humor, and have the confidence to voluntarily reveal more personal aspects to them. The \textit{CFS} is proposed as consisting of positive relationships of the adolescent with their peers and positive links with their school in terms of a sense of \textit{Bel} and \textit{Sup}, clarity of the rules and values of the school, and the provision of educational and recreational activities by the school.
In general, it has been shown by information gain values that the \textit{AfC} factor contributes the most to \textit{PP}, \textit{Bon} contribute the most to \textit{CFS}, and \textit{Bel} contributes the most to \textit{PYD}. Furthermore, it is found that positive relationships with parents have a greater impact on the adolescent development and well-being than do school bonds, \citet{arslan_2022, balaguer_2021} or the school climate, \citet{balaguer_2021, oliva_2011}. 
However, one of the main contributions of this work is that the conditional probability of the \textit{PP} and \textit{CFS} quintiles for the \textit{PYD} can be verified, thanks to the analysis provided by the BNM, particularly the contour figures. Thus, it is shown that a low \textit{CFS} perception negatively affects \textit{PYD}, independently of the parenting perception. A negative \textit{PYD} may be had by the adolescent with a positive \textit{PP} if their school context is negative. Therefore, \textit{Bel} is identified as an essential mechanism that contributes to understanding the effect of positive, strengths-based parenting on the development and well-being of youth, \citet{arslan_2022}. For a medium \textit{PYD} level, it is observed that \textit{PP} begins to become independent from \textit{CFS}. Indeed, the family is recognized as the main agent of socialization and positive development. Family involvement is seen to facilitate adolescent adjustment and school identification, as well as bonds with their teachers and peers, which contributes to the youth’s perception of acceptance and support from the school, \citet{law_2013, uslu_2017}. Precisely, to reach high or very high \textit{PYD} levels, the adolescent needs both high and very high \textit{PP} and \textit{CFS} levels, especially the latter. The adolescent requires support and bonds with school as well as positive relationships with classmates, in addition to family support characterized by frequent closeness and affection from parents, \citet{animosa_2018}.

In this way, it is shown that the context plays a determining role in the interaction with personal adolescent development, \citet{delafuente_2017}. In particular, a positive school context, which enables the adolescent to perceive school support and bonds, promotes their development at multiple levels: social, emotional, academic, and health, \citet{waters_2009, wong_2021}. An adolescent cannot develop positively with low \textit{CFS}, regardless of the perceived quality of parenting. However, excellent positive parenting greatly enhances the effect of the school context for optimal positive personal development. Hence, the importance of schools actively involving parents to foster their children's school bonds, \citet{arslan_2022}.

In conclusion, this article argues for the necessity of reinstating theorists in disciplines to their role in defining the structure of models. This role has attempted to be supplanted by machines, which have repeatedly shown that they cannot operate alone by merely hoping that the exponential increase in information will be sufficient to generate the needed knowledge. It is the collaborative work of causal theoretical models, computational algorithms, and statistical methods that should be proposed as a working methodology for advancing knowledge a synergistic rather than competitive effort between these approaches.

\section*{Declarations}

\subsection*{Funding}
Not applicable.

\subsection*{Conflicts of interest/Competing interests}
Not applicable.

\subsection*{Ethics approval}
Not applicable.

\subsection*{Consent to participate}
Not applicable.

\subsection*{Consent for publication}
Not applicable.

\subsection*{Availability of data and materials}
The data used in this study are openly available in a GitHub repository at \url{https://github.com/ebenitezs/SEM_BN}, \citet{benitez_2024}. For reproducibility and further analysis, all materials can be accessed and downloaded.

\subsection*{Code availability}
Not applicable.
The R code used in this study are openly available in a GitHub repository at \url{https://github.com/ebenitezs/SEM_BN}, \citet{benitez_2024}. For reproducibility and further analysis, all materials can be accessed and downloaded.

\subsection*{Authors' contributions}
A.B. was responsible for the theoretical model and the interpretation of the results in light of the analysis. E.B. contributed the original idea, developed the code, and drafted the initial versions of the manuscript. Both authors collaboratively wrote the final document.

\clearpage 

\bibliography{sn-bibliography} % Nombre del archivo .bib sin la extensión

\end{document}